\documentclass{tlp}
\usepackage{times}
\usepackage{epsf,epsfig,subfigure,latexsym,makeidx,latexsym,xspace,amssymb,abbrev}
\usepackage{graphicx}           
\usepackage{amsmath}
\usepackage{color}
\usepackage{dashrule}

\usepackage{float}
\usepackage{url}
\usepackage{verbatim}
\usepackage{latexsym}
\usepackage{epic}
\usepackage{ecltree}

\newcommand{\bc}{\begin{center}}
\newcommand{\ec}{\end{center}}

\newcommand{\longline}{\noindent\rule{\textwidth}{.01in}}

\newtheorem{dummylemma}{Dummy}[section]
\newenvironment{example}
               {\begin{ExampleN}~\rm} {\hfill  $\Box$\end{ExampleN}}
\newtheorem{ExampleN}{\bf Example}[section]

\newtheorem{JDefinitioN}[section]{\bf Definition}

  {\begin{list}{} %
	{\setlength{\leftmargin}{10pt}%
         \setlength{\rightmargin}{10pt}%
	}%
  }%
  {\end{list}}

\floatstyle{ruled}
\newfloat{Algorithm}{thpb}{lop}[section]

\newtheorem{DefinitioN}[dummylemma]{\bf Definition}

\newcommand{\callgr}{IDG}
\newcommand{\Callgr}{IDG}

\newcommand{\cD}{{\cal D}}

\newcommand{\version}{Version 1(beta)}

\newcommand{\bi}{\begin{itemize}}
\newcommand{\ei}{\end{itemize}}
\newcommand{\newcpinst}{{\sf preservedViewMember}}
\newcommand{\oldinc}{manual incremental tabling}
\newcommand{\newinc}{automatic incremental tabling}
\newcommand{\Newinc}{Automatic incremental tabling}
\newcommand{\newtab}{pervasive tabling}

\newcommand{\abstraction}{IDG abstraction}

\begin{document}

\author{Terrance Swift\\
         NOVALincs, Universidade Nova de Lisboa\\
         \email{terranceswift@gmail.com}}

\title{Incremental Tabling in Support of Knowledge Representation and Reasoning}

\date{}

\sloppy
\maketitle

\begin{abstract}
Resolution-based Knowledge Representation and Reasoning (KRR) systems,
such as Flora-2, Silk or Ergo, can scale to tens or hundreds of
millions of facts, while supporting reasoning that includes Hilog,
inheritance, defeasibility theories, and equality theories.  These
systems handle the termination and complexity issues that arise from
the use of these features by a heavy use of tabled resolution.  In
fact, such systems table by default all rules defined by users, unless
they are simple facts.

Performing dynamic updates within such systems is nearly impossible
unless the tables themselves can be made to react to changes.
Incremental tabling as first implemented in XSB~\cite{Saha06}
partially addressed this problem, but the implementation was limited
in scope and not always easy to use.  In this paper, we introduce {\em
  \newinc{}} which at the semantic level supports updates in the
3-valued well-founded semantics, while guaranteeing full consistency
of all tabled queries.  \Newinc{} also has significant performance
improvements over previous implementations, including lazy
recomputation, and control over the dependency structures used to
determine how tables are updated.

\end{abstract}


\section{Introduction}
Tabled Logic Programming has supported a variety of applications that
would be difficult to implement in Prolog alone, including model
checking, program analysis, ontology-based deductions and decision
making for collaborative agents.  Typically such applications are
written mainly as Prolog programs, but with a subset of the predicates
tabled in order to support termination, reduce complexity, to use
well-founded negation or to exploit other features.

However, systems such as Flora-2~\cite{YKHZ13} and its extensions:
Silk (cf. {\sf silk.semwebcentral.org}), Ergo (cf. {\sf
  coherentknowledge.com/publications}) and the RAVE system (cf. {\sf
  www.sri.com/about/people/grit-denker}) have been recently developed
for knowledge representation and reasoning (KRR), and rely on tabled
resolution for their computational underpinning.  For instance,
Flora-2~\cite{YKHZ13}, which is based on XSB~\cite{SwiW12}, supports
the non-monotonic inheritance of F-logic, prioritized defeasibility
with multiple levels of conflicts, rule identifiers, function symbols,
logical constraints, and HiLog.
Silk and Ergo, both based on Flora-2, support all of the above
features plus {\em omni axioms}, which are contrapositional rules
whose bodies and heads are comprised of any formulas that can be
supported by the Lloyd-Topor transformation~\cite{LlT84}.

As an example of using these features,
given the sentence: {\em A contractile vacuole is inactive in an
  isotonic environment} from~\cite{Camp10}, a tool called Linguist
({\sf www.haleyai.com}) produces a Silk or Ergo formula in a mostly
automatic manner (knowledge engineers may have to choose between
translations in ambiguous cases), resulting in the axiom:
%
\begin{small}
{\em
\begin{tabbing}
fooofoo\=foofoofoofoofoo\=fooooooooo\=oooooooooooooooooo\=ooooooooooooo\=\kill
\>     forall(?x6)\verb|^|contractile(vacuole)(?x6)  \\
\> \>         ==$>$ forall(?x9)\verb|^|isotonic(environment)(?x9) \\
\> \>         ==$>$ inactive(in(?x9))(?x6);
\end{tabbing}
}
\end{small}
\noindent
Such an axiom is next translated into several Flora-2 rules about
conditions of contractile vacuoles, inactive contractile vacuoles, and
isotonic environments.  These Flora-2 rules are then transformed to
support HiLog, defeasibility and other features, resulting in numerous
normal rules executed in XSB.  Once a knowledge base has been
constructed from axioms such as the one above, queries can be made
such as: {\em If a Paramecium swims from a hypotonic environment to an
  isotonic environment, will its contractile vacuole become more
  active?}  The translation of queries is similar to that of
knowledge, but may include {\em hypothetical} information, e.g., that
{\em ?x} is a Paramecium swimming from a hypotonic environment to an
isotonic environment.
Knowledge bases themselves are built from a collection of rules and
omni axioms usually written by different knowledge engineers using a
shared background vocabulary.  The limited coordination among
knowledge engineers is critical for producing knowledge bases at a low
cost.  

All of the the KRR-systems mentioned above employ what may be called
{\em \newtab} where a predicate is tabled unless it is explicitly
declared non-tabled.
Such programs have an operational behavior that is vastly different
from (tabled) Prolog.  Among other matters, as many of these tables
represent background knowledge, it is critical for good system
performance to reuse tables between queries.  However, because queries
may include hypothetical knowledge, and because knowledge bases are
created by interactively adding or modifying rules, good performance
demands the use of {\em incremental tabling}~\cite{SaRa05,Saha06}.
%
%

The main idea behind incremental tabling is to maintain an {\em
  Incremental Dependency Graph (IDG)}, indicating how tabled goals
depend both on dynamic code and on one another.  When an update is
made to dynamic code, the \callgr{} is traversed, and affected tables
are updated if necessary.  However, while previous versions of
incremental tabling were robust enough to support a commercial
application~\cite{dss-ker}, they were not sufficient to support
high-level KRR applications.  Most significantly, a programmer had to
decide when tables were updated: either an update was forced
immediately upon an assert or retract, or the programmer performed
``bulk'' updates, after which a command propagated the updates to all
affected tables.  This methodology was complicated and had semantic
drawbacks: unless an update was manually invoked, there was no
guarantee that tables would be updated and no provision for stronger
forms of view consistency.  In fact, because of the brittleness caused
by the need for low-level control along with other drawbacks, previous
versions of incremental tabling, (designated here as {\em \oldinc{}})
were suitable only for careful use by tabling experts.

Support for pervasive tabling requires that a tabling engine be
redesigned in several ways, including the mechanisms whereby tables
are updated.  This paper introduces {\em \newinc{}} to support
applications that rely on \newtab{} such as the KRR-systems described
above.  The papers major contributions are:
\bi
\item A description of core changes that allow table updates to be
  made in a safe and efficient manner: first, tables are updated
  automatically and efficiently by {\em lazy recomputation}; second,
  updates always guarantee view consistency for incremental tables.

\item A description of how incremental recomputation is extended to
  support updates according to the three-valued well-founded
  semantics.
\item Introduction of the notion of {\em \abstraction} to reduce the
  size of the \callgr{} when necessary.

%
\item Detailed performance analyses of \newinc{} for both small
  program fragments and for KRR-style examples over Extensional
  Databases (EDBs) up to size $\cO(10^7)$.  These results indicate
  that \newinc{} efficiently supports the
  KRR uses previously mentioned, and may also provide a basis for {\em
    reactive KRR}.
\end{itemize}
\Newinc{} is available in the current version of XSB.  In addition to
the extensions mentioned above, its implementation is based on a
significant rewriting
of the previous implementation of \oldinc{}.  Incremental tabling is
not yet available in tabling engines other than XSB.  However, while
transparent incremental tabling adds data structures such as the IDG,
it interfaces with a tabling engine mostly through routines for
maintaining table space.  Accordingly, most of the features described
below are relatively portable, as tabling engines have similar table
space operations, and sometimes similar data structures.

\vspace{-.15in}

\section{A Review of Manual Incremental Tabling} \label{sec:background}
In this section we describe the previous version of incremental
tabling using the main data structures and algorithms
of~\cite{Saha06}, which form the starting point for the features of
\newinc{} described in later sections.
The description is as self-contained as possible, but sometimes uses
the terminology of the SLG-WAM~\cite{SaSw98}.

\begin{figure}
\begin{minipage}[b]{0.45\linewidth}
\begin{footnotesize}
{\em
\begin{tabbing}
fooofoo\=foofoofoofoofoo\=fooooooooo\=oooooooooooooooooo\=ooooooooooooo\=\kill
:- table t\_1/1, t\_2/1, t\_4/1, t\_5/1 as incremental. \ \ \ \ \ \\\
t\_1(X)\> :- t\_4(X),tnot(t\_2(X)).\\
t\_4(X)\>:- t\_5(X). \>      t\_4(X):- t\_4(Y),t\_5(X).\\
t\_5(X)\>:- nt\_1(X).    \> t\_2(X):- q(X).\\
nt\_1(X)\>:- p(f(X)).\>     nt\_1(X):- p(g(X)).\\
\> \\
:- dynamic p/1, q/1 as incremental. \\
p(f(1)).\>\>                q(1).
\end{tabbing}
}
\end{footnotesize}
\end{minipage}
\begin{minipage}[b]{0.45\linewidth}
\includegraphics[width=0.6\textwidth]{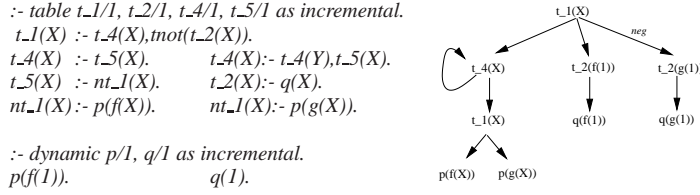}
\end{minipage}
\caption{A Program $P_{inc}$, and schematic Incremental Dependency
  Graph (IDG) for the query {\em t\_1(X)}} \label{fig:p-inc}
\end{figure}

Fig.~\ref{fig:p-inc} shows an XSB program $P_{inc}$ where predicates
are declared to use incremental tabling.  In general both tables and
dynamic code may be declared with various attributes: not only {\em
  incremental} as here, but also {\em subsumptive}, {\em
  trie-indexed}, and so on.
Note that {\em tnot/1} is an XSB operator for tabled negation.
Execution of the query {\em t\_1(X)} creates the {\em Incremental Dependency Graph (IDG)} schematically
shown in Fig.~\ref{fig:p-inc}.
The \callgr{} has a node for each tabled subgoal but not for
non-tabled subgoals such as {\em nt\_1(X)} -- though the
bindings made by the rules for {\em nt\_1/1} are implicitly propagated.
Leaf nodes in the \callgr{} correspond to predicates such as {\em p/1}
and {\em q/1} that are declared to be both dynamic and incremental.
Each downward edge in a \callgr{} represents an element of the {\em
  direct dependency} relation; the inverse relation is the {\em direct
  affected} relation.  Note that paths in the \callgr{} may be cyclic.
%

At the level of data structures, each node in the \callgr{} is
represented via an {\em IDG node frame} (Fig.~\ref{fig:callnode}).
For a tabled incremental subgoal {\em t/n}, the IDG node frame is
created by the {\sf tabletry} instruction, by registering it into the
subgoal trie for {\em t/n}~\footnote{In XSB, the default data
  structure for tabled subgoals and their answers is based on
  tries~\cite{RRSSW98}.  While XSB offers basic support for answers
  that are ``hash-consed''~\cite{ZhoH12} and not maintained as tries,
  our presentation assumes subgoal and answer tries throughout.}, and linking it
with the {\em subgoal frame}, which contains information about each
tabled subgoal.
For dynamic incremental subgoals a new SLG-WAM instruction, {\sf
  try\_dynamic\_incremental} performs these tasks.
Each time a (tabled or dynamic) incremental subgoal $S$ is called, the
\callgr{} may be updated. If $S$ is new, an IDG node frame is created;
also whether or not $S$ is new, if $S$ has a nearest tabled subgoal
$S_{par}$ as an ancestor, edges between $S$ and $S_{par}$ are added if
not already present.  As answers are derived for $S$, their count is
maintained in the {\em nbr\_of\_answers} field of the IDG node frame.

{\small
\begin{figure}[hbtp] 
\centering
\begin{tabular}{l|l|l} \cline{2-2}
  & {\em affected\_edges} & Subgoals that this subgoal directly affects\\ \cline{2-2}
  & {\em dependent\_edges} & Subgoals upon which this subgoal directly depends\\ \cline{2-2}
  & {\em subgoal\_frame} & Pointer back to the subgoal frame\\ \cline{2-2}
  & {\em nbr\_of\_answers} & Counts the number of answers rederived   \\ \cline{2-2}
  & {\em previous\_IDG\_node} & Used to determine if re-evaluation has changed the set of answers  \\ \cline{2-2}
  & {\em new\_answer} & set to true if a new answer has been derived\\ \cline{2-2}
  & {\em falsecount}  & determines whether subgoal is valid\\ \cline{2-2}
\end{tabular}
\caption{The IDG node frame for incremental tables}\label{fig:callnode}
\end{figure}
}
At a high level, the use of the \callgr{} is easy to understand.  If a
fact, say {\em p(g(2))}, is asserted, the incremental update subsystem
must call {\sf traverse\_affected\_nodes()}
(Fig.~\ref{alg:basic-update}) to traverse the \callgr{}.  Separate
traversals start from each leaf node with which {\em p(g(2))} unifies,
and the traversals will increment the {\em falsecount} field of their
IDG node frame (cf. Fig.~\ref{fig:callnode}), marking them as {\em
  invalid} (i.e., having a {\em falsecount} greater than 0).  As it is
unclear whether sensible semantics can be given to updating a subgoal
that is incomplete (i.e., that is still being computed), a permission
error is thrown if this is attempted.  In our running example,
assuming that no nodes in the \callgr{} are already invalid, the
algorithm will traverse depth-first through all nodes affected by {\em
  p(g(X))} (directly or indirectly).  In so doing, the affected
non-leaf nodes are added to a global {\em invalid list} in the same
order.  In our example, the nodes for {\em t\_5(X)}, {\em t\_4(X)} and
{\em t\_1(X)} are traversed, and the invalid list represents this
sequence.

Several properties of the traversal are worth noting.  First, use of
the {\em falsecount} field in {\sf traverse\_affected\_nodes()}
prevents the same node from being traversed multiple times.  Also,
note that invalidation simply represents {\em some} change in the
underlying data so that retracts are handled in the same manner as
asserts, and both positive and negative dependencies are treated in
the same way.  In fact, since the traversal starts with dependency
leaf nodes that unify with a given atom, propagation of a rule update
is handled in the same manner as a fact update: {\sf
  traverse\_affected\_nodes()} is invoked for leaf nodes that unify
with the rule head.  In either case, the unification of leaf nodes
with a given atom can also prevent unnecessary updates: for instance,
if the fact {\em q(g(2))} were added, it would not cause any update,
since no leaf node of the \callgr{} unifies with this fact.

\begin{figure}[btp] 
{\sf
\begin{tabbing}
fooo\=foo\=foo\=foo\=foo\=foo\=fooooooooooooooooooooooooooooo\=ooooooooooooo\=\kill
\underline{traverse\_affected\_nodes(IDG node frame {\em IDGN})} \\ 
\> {\rm /* {\em IDGN} is the IDG node frame for an incrementally tabled predicate */} \\ 
\> If the table associated with {\em IDGN} is not completed, throw a permission exception \\ 
\> For each {\em IDGN}$_{aff}$ that is directly affected by {\em IDGN} \\ 
\> \> {\em  IDGN}$_{aff}.falsecount$++; \\ 
\> \> If ({\em  IDGN}$_{aff}.falsecount$ == 1) traverse\_affected\_nodes({\em IDGN}$_{aff}$) \\ 
\> \> Add {\em IDGN} to the global invalid list.\\ 
\\ 
\underline{incremental\_reeval(IDG node frame {\em IDGN})}  \> \> \> \> \> \> \>
    {\rm /* $S$ is the subgoal to be recomputed */}\\
\> If {\em IDGN}$.falsecount > 0$ \\ 
\> \> Let $S_T$ be the subgoal frame associated with {\em IDGN} (i.e., $S_T = ${\em IDGN.subgoal\_frame}) \\ 
\> \> For each $A$ in $S_T.answer\_list$ \\ 
\> \> \> Mark $A$ as deleted, but do not adjust answer trie choice points or reclaim space \\ 
\> \> Create a new IDG node {\em IDGN}$_{new}$ for $S_T$ \\ 
\> \> {\em IDGN}$_{new}.new\_answer := false$; {\em IDGN}$_{new}.falsecount$ = {\em IDGN}$_{new}.nbr\_of\_answers$ = 0 \\
\> \> Call  $S$ and for each new derived answer $A_{deriv}$\\
\> \> \>  Increment {\em IDGN}$_{new}.nbr\_of\_answers$\\
\> \> \>  If $A_{deriv}$ was marked as deleted, remove the deletion mark\\
\> \> \>  Else {\em IDGN}$_{new}.new\_answer$ = true \\
\> \> After completion of $S$, for each $A$ in $S_T.answer\_list$ \\
\> \> \> If $A$ is still marked as deleted, remove $A$ from $S_T.answer\_list$ \\
\> \> \> \> Reset answer trie choice points and reclaim space for $A$ \\
\> \> If {\em IDGN}$_{new}.new\_answer = false$ and {\em IDGN}$_{new}.nbr\_of\_answers$  = {\em IDGN}$.nbr\_of\_answers$ \\
\> \> \> propagate\_validity(IDG node frame {\em IDGN}) \\
\>  \\
\underline{propagate\_validity(IDG node frame {\em IDGN})} \\
\> \> For each {\em IDGN}$_{aff}$ that is directly affected by {\em IDGN} \\
\> \> \> {\em IDGN}$_{aff}.falsecount$- - \\
\> \> \> if {\em IDGN}$_{aff}.falsecount$ == 0 propagate\_validity({\em IDGN}$_{aff}$)
\end{tabbing}
}
\caption{Schematic algorithms for \oldinc} \label{alg:basic-update}

\end{figure}

After the invalidation phase is finished, reevaluation of the affected
nodes may be done either immediately, or at a later time through an
explicit command.  Note that once the invalid list has been set up,
the affected tables can be updated in a bottom-up manner simply by
removing them in order from the list.  Specifically, for each IDG node
{\em IDGN} removed from the invalid list, {\sf
  incremental\_reeval({\em IDGN})} called
(Fig.~\ref{alg:basic-update}).  If {\em IDGN}$.falsecount$ is 0, the
subgoal does not need to be recomputed.  Otherwise, the answers for
$T$, the table associated with {\em IDGN}, are marked as deleted,
although their space is not reclaimed~\footnote{The {\em answer list}
  of an answer trie, which allows easy traversal of all answers in the
  trie, is reclaimed at the completion of each non-incremental
  table, but retained by incremental tables for traversals during
  re-evaluation.}.  A new IDG node {\em IDGN}$_{new}$ is created for
$T$, and its {\em previous\_IDG\_node} field is set to the old IDG
node, {\em IDGN} (cf. Fig.~\ref{fig:callnode}).  The subgoal for $T$
is re-evaluated, and for each answer $A$, {\em
  IDGN}$_{new}.nbr\_of\_answers$ is incremented; in addition if $A$ is
new, (i.e., the addition of the answer $A$ does not undelete a
previously obtained answer) {\em IDGN}$_{new}.new\_answer$ is
incremented.  Clearly, if {\em IDGN}$.nbr\_of\_answers$ is not equal
to {\em IDGN}$_{new}.nbr\_of\_answers$, the answers for $T$ have
changed; also if the two numbers are the same but {\em
  IDGN}$_{new}.new\_answer$ is set, the answers for $T$ have changed.
Otherwise, the answers for $T$ have not changed, and the subgoals $T$
affects are traversed to decrement their $falsecount$ fields, which
may transitively prevent other subgoals from having to be recomputed
(cf.  {\sf propagate\_validity()} in Fig.~\ref{alg:basic-update}).
\vspace{-.15in}

\section{Supporting Well-Founded Negation} \label{sec:wfs}

A necessary extension for incremental tabling to support KRR
applications of the type mentioned in the introduction is to support
full well-founded negation.  KRR applications make use of the {\em
  undefined} truth value to represent conflicts in the defeasibility
theory used by a program, as well as to handle infinite models through
a type ofanswer abstraction called {\em restraint}~\cite{GroS13}, and
to support debugging of KRR programs (cf.~\cite{Nais06}).

As mentioned in the previous section, the \callgr{} maintains
information about the dependency and affected relations without
representing whether these changes are positive or negative.  One
advantage of this is that \oldinc{} is correct for stratified negation
--- here, meaning well-founded negation with two-valued models.
However, to support full well-founded negation, the update process
must handle tables in which some atoms are {\em undefined}.  To
explain how this is done, we overview those aspects of well-founded
negation in SLG resolution~\cite{CheW96} that are relevant to the
incremental update algorithms.

Essentially, a query evaluation by SLG resolution builds up a partial
model of those parts of a program $P$ that are relevant to the query.
To make this specific, an SLG evaluation $\cE$ is modeled as a
sequence of states, called {\em forests}.  Let $\cF$ be one such
forest in $\cE$.  $\cF$ contains a set of tabled subgoals that have
been encountered so far in $\cE$.  Each such tabled subgoal $S$ in
$\cF$ is associated with a table $T_S$ containing computed answers for
$S$; $T_S$ may be marked as {\em completed} in $\cF$ if it has been
determined that all necessary 
resolution has been performed to derive answers for $S$.  To support
3-valued interpretations of $\cF$, answers are distinguished as {\em
  unconditional} answers representing true derivations, and {\em
  conditional} answers representing derivations of atoms with truth
value {\em undefined}.  Accordingly, let $T_S$ be a completed table
for a subgoal $S$ in $\cF$, and let $S_G$ be an atom in the ground
instantiation of $S$.  $S_G$ is true if it is in the ground
instantiation of some unconditional answer in $T_{S}$, and is false if
is {\em not} in the ground instantiation of {\em any} answer in
$T_{S}$ (conditional or unconditional).

Formally, for a subgoal $S$, a conditional answer has the form 
$S\theta \mif{} DL|$ where 
$S\theta$ is termed the {\em answer substitution}; and $DL$, the {\em
  delay list}, is a list of literals needed to prove $S\theta$ but
whose resolution has been delayed because they do not have a
well-founded derivation (based on the current state of the evaluation
if $T_S$ is not completed).  During the course of an evaluation, if a
literal $L$ in a delay list becomes {\em true} or {\em false}, the SLG
{\sc simplification} operation respectively removes $L$ from the delay
list or indicates that the conditional answer itself is false.

\begin{example} \label{ex:cond}
The goal {\em p(X)} to the program 
\begin{tabbing}
fooofoooooooooo\=foofooooooooooooo\=foofooooooooooooo\=foo\=foo\=fooooooooooooooooooooooooooooo\=ooooooooooooo\=\kill

{\em p(1)} \> {\em     p(2):- not q(2)} \>  {\em p(2):- not q(3)} \>    {\em q(X):- not p(X)}

\end{tabbing}
has an unconditional answer {\em p(1)} along with two conditional
answer: {\em p(2):- not q(2)$|$}, {\em p(2):- not q(3)$|$}.  Note that
the delay lists for answers to {\em p(2)} contain only {\em undefined}
literals upon which {\em p(2)} {\em directly} depends (e.g., {\em not
  q(2), not q(3)}), but not indirect dependencies such as {\em not
  p(3)}.
\end{example}

As mentioned above, XSB represents both tabled subgoals and their
answers using tries, a representation that is supported by other
Prologs such as YAP~\cite{SCDR12} and Ciao~\cite{HBCLMMP12}.  In XSB
this representation is extended as follows~\cite{SaSW00}.  If an atom
$A_{cond}$ is {\em undefined}, the leaf node of the answer
representing $A_{cond}$ points to an {\em answer information frame}
which in turn points to other answers conditional on $A_{cond}$ as
well to a {\em delay trie} representing all delay lists upon which
$A_{cond}$ is conditional.  In Example~\ref{ex:cond} the delay trie
for {\em p(2)} would contain the lists {\em [not q(2)]} and {\em [not
    q(3)]}.  Whenever an unconditional answer $S\theta$ is derived in
a table for subgoal $S$, the answer information frame and delay trie
for conditional answers to $S\theta$ are deallocated if they exist.

To extend incremental recomputation to correctly handle changes
involving conditional answers, several previously unconsidered cases
must be addressed for a given answer substitution $S\theta$ in a table
$S$.  Each case below considers only those answers in
the table $S$.
\begin{itemize}
\item {\em Informational Weakening 1,} There were previously no answers
  for $S\theta$; after the update there are one or more conditional
  answers for $S\theta$.
\item {\em Informational Weakening 2,} There was previously an
  unconditional answer for $S\theta$; after the update there are one
  or more conditional answers for $S\theta$.
\item {\em No Informational Change,} There were previously one or more
  conditional answers for $S\theta$; after the update further
  conditional answers $S\theta$ were added, or some but not all
  conditional answers for $S\theta$ were deleted.
\item {\em Informational Strengthening 1,} There were previously one or
  more conditional answers for $S\theta$; after the update $S\theta$
  becomes {\em true}, with an unconditional answer.
\item {\em Informational Strengthening 2,} There were previously one or
  more conditional answers for $S\theta$; after the update $S\theta$
  becomes {\em false}, with no answers.
\end{itemize}

The cases above are grouped by their action on the information
ordering of truth values, where both {\em true} and {\em false} are
stronger than {\em undefined}.
From the perspective of table updates, no action need be taken in the
case of {\em No Informational Change}, as the truth value of $S\theta$
is unchanged.  To see this, recall that delay lists contain only
direct dependencies.  Thus any answer $A'$ that is conditional on
$S\theta$ will contain $S\theta$ or $not\ S\theta$ in its delay lists
so that changes to the delay list of $S\theta$ need not be propagated.
Strengthening and weakening of answers are addressed by the extensions
to {\sf incremental\_reeval()} shown underlined in
Fig.~\ref{alg:wfs-update}.
\begin{figure}[hbtp] 
{\sf
\begin{footnotesize}
\begin{tabbing}
fooo\=foo\=foo\=foo\=foo\=foo\=fooooooooooooooooooooooooooooo\=ooooooooooooo\=\kill
{\sf incremental\_reeval(IDG node frame {\em IDGN})} \> \> \> \> \> \> \>
    {\rm /* $S$ is the subgoal to be re-computed */}\\
\> If {\em IDGN}$.falsecount > 0$ \\ 
\> \> Let $S_T$ be the subgoal frame associated with {\em IDGN} (i.e., $S_T = ${\em IDGN.subgoal\_frame}) \\
 \> \> if $S_T.occp\_num > 0$ preserve\_occp\_views($S_T$) 
\> \> \> \> \> {\rm /* Ensure view consistency: see Section~\ref{sec:consistency} */}\\ 
5\> \> For each answer $S\theta$ in $S_T.answer\_list$ \\
 \> \> \> $S\theta.deleted = true$ \\
\> \> \> \underline{If $S\theta$ is unconditional $S\theta.unconditional = true$ else $S\theta.unconditional = false$ } \\
\> \> Create a new IDG node {\em IDGN}$_{new}$ for $S_T$ \\ 
\> \> {\em IDGN}$_{new}.new\_answer := false$; {\em IDGN}$_{new}.falsecount$ = {\em IDGN}$_{new}.nbr\_of\_answers$ = 0 \\
10\> \> Call $S$ and for each new derived answer $S\theta \mif{} DL|$ \> \> \> \> \> \\
\> \> \> If $S\theta.deleted == true$ \\
\> \> \> \>  $S\theta.deleted = false$; {\em IDGN}$_{new}.nbr\_of\_answers++$\\
 \> \> \>  \underline{If $S\theta.unconditional == false$ but $S\theta$ is now unconditional} \\
\> \> \> \> \underline{$S\theta.unconditional = true$; invoke simplification}\\
15\> \> \>  Else {\rm /* $S\theta$.deleted was false */} {\em IDGN}$_{new}.new\_answer$ = true \\
\> \> After completion of $S$, for each $S\theta$ in $S_T.answer\_list$ \> \> \> \> \> \\
\> \> \> If $S\theta.deleted == true$, remove $S\theta$ from $S_T.answer\_list$ \\
\> \> \> \> \underline{If $S\theta.unconditional = false$ invoke simplification} \\
\> \> \> \> Adjust trie choice points and reclaim space for $S\theta$ \\
20\> \> \> \underline{Else if $S\theta.unconditional == true$, and $S\theta$ is now conditional}\\
 \> \> \> \> \underline{{\em IDGN}$_{new}.new\_answer = true$} \\
\> \> {\em IDGN.reeval\_ready = compute\_dependencies\_first} \\
 \> \> If {\em IDGN}$_{new}.new\_answer == false$ and {\em IDGN}$_{new}.nbr\_of\_answers  = ${\em IDGN}$.nbr\_of\_answers$ \\
\> \> \> propagate\_validity({\em IDGN})
\end{tabbing}
\end{footnotesize}
}
\caption{Schematic algorithm for updates in \newinc{}} \label{alg:wfs-update}
\end{figure}

As shown in Fig.~\ref{alg:wfs-update} setup for the re-derivation of
$S$ now also sets a new {\em unconditional} field of an answer,
representing whether the answer was unconditional at the start of the
re-derivation (line 7).  In the re-derivation, {\em
  IDGN}$_{new}.nbr\_of\_answers$ is incremented whenever a new answer
substitution $S\theta$ is encountered, whether $S\theta$ is
conditional or unconditional (Fig.~\ref{alg:wfs-update} lines
11-12), so that {\em IDGN}$_{new}.nbr\_of\_answers$ will be updated at
most once regardless of how many conditional answers exist for
$S\theta$.  Thus, there are no changes required for {\em Informational
  Weakening 1} as the addition of new conditional and unconditional
answer substitutions is handled in the same manner.  Also, if more
than one conditional answer is derived for $S\theta$, only the first
will increment {\em IDGN}$_{new}.nbr\_of\_answers$, in effect handling
the case of {\em No Informational Change}.  A similar check of
$S\theta.unconditional$ during re-derivation (lines 13-14) handles
{\em Informational Strengthening 1}, the case where $S\theta$ had only
conditional answers, but is now unconditional.  This case can actually
be handled directly by SLG {\sf simplification} and does not require
propagation through the incremental update system.  Once $S$ has been
rederived, its answer list is traversed as before
(cf. Fig.~\ref{alg:basic-update}).  During this traversal, line 18
handles the case of {\em Informational Strengthening 2} where
$S\theta$ had been conditional but is now false and uses {\sf
  simplification}; Lines 20-21 handle {\em Informational Weakening 2}
where $S\theta$ had been {\em false} but now is {\em undefined}.

Fig.~\ref{alg:wfs-update} reflects a bilattice of
the information ordering and the truth ordering (where $true >$ {\em
  undefined}$ > ${\em false}).  As discussed, SLG {\sf simplification}
propagates changes of an answer's truth value when it is
informationally strengthened.  Changes to the truth value of $S\theta$
that reflect a strengthening in the truth ordering can be detected
during re-derivation ({\em Information Weakening 1}, {\em Information
  Strengthening 1}).  Changes that reflect a weakening in the truth
ordering must wait until the re-derivation is complete ({\em
  Information Weakening 2}, {\em Information Strengthening 2}).

\mycomment{
Thus case 1, where there were previously no answers for $S\theta$ is
handled in the same way for both conditional and unconditional
answers.  Case 2, where there was previously an unconditional answer
for $S\theta$ but there is now only conditional answer(s) for $S$ is
handled by checking a new bit field, $unconditional$, added to trie
nodes.  During the traversal of the answer list for $S$ prior to its
re-derivation (Fig.~\ref{alg:wfs-update} lines 4-6) a bit in the
leaf trie node of each unconditional answer is set to true.  Later, in
the answer list traversal after the re-derivation (lines 13-18) the
$unconditional$ bit is checked to determine that an unconditional

Fig.~\ref{alg:wfs-update} also indicates how cases 4 and 5 are
handled.  Case 4, an answer substitution that was previously
conditional and is now unconditional can be caught as soon as the
unconditional answer is derived (line 12).  Case 5, an answer
substitution that was previously conditional and is now false, cannot
be caught until the post-reevaluation traversal (line 15).  In both
cases, SLG simplification is invoked, making use of fine-grained
dependency pointers among answers and subgoals~\cite{SaSW00}.  Thus,
cases 4 and 5 are handled in a more ``incremental'' manner than the
other cases: the truth values of answers are updated directly, without
the need of further re-evaluation of affected subgoals.
}
\vspace{-.15in}

\section{Ensuring Transparency through Lazy Recomputation and View Consistency} \label{sec:lazy}

Perhaps the main drawback of \oldinc{} is the level of control it
requires from a programmer.  A programmer can specify that an
incremental update is to be done immediately after an assert or
retract, but this is inefficient when multiple updates are required.
Alternatively, a programmer can specify that an assert or retract
simply invalidate affected subgoals, but later must make a call to
reevaluate subgoals on the invalid list.  In either case, if choice
points exist to an incrementally tabled subgoal $S$ that is completed,
the semantics of an update are undefined (and in fact the program may
crash).  In addition to these issues, \oldinc{} may cause unnecessary
work as all affected goals are recomputed even if they are never
re-queried.  We show how these problems are fixed in \newinc.

\subsection{Lazy Recomputation}
In lazy recomputation assert and retract hooks invalidate tables when
a change is made to a dynamic incremental predicate.  However, an
invalid subgoal $S$ is not re-evaluated until it is called, at which
time incremental tabled subgoals upon which $S$ depends are also
re-evaluated.  The algorithm for lazy recomputation is shown in
Fig.~\ref{fig:lazy} within a schematic description of the SLG-WAM's
{\sf tabletry} instruction, which is executed upon calling a tabled
subgoal~\footnote{XSB's {\sf tabletry} instruction is substantially
  more complex as it supports call subsumption, subgoal abstraction,
  multi-threaded tabling and other features.}.
Specifically, if $S$ is completed and invalid, lazy recomputation is
handled within lines 12-17, using a {\em reeval\_ready} field, which
\newinc{} adds to each IDG node frame.  If the {\em
  reeval\_ready} field for $S$ is set to {\em
  compute\_dependencies\_first} the IDG nodes upon which $S$ depends
are traversed in a depth-first manner by {\sf
  traverse\_dependent\_nodes()} and the traversed subgoals are added
to the invalid list (Fig.~\ref{fig:lazy}).  This predicate, analogous
to {\sf traverse\_affected\_nodes()} of Fig.~\ref{alg:basic-update},
traverses dependency edges rather than affected edges.
 Once the invalid list is constructed, its subgoals are recomputed by
 {\sf recompute\_dependent\_tables()} which iteratively calls the
 version of {\sf incremental\_reeval()} in Fig.~\ref{alg:wfs-update}.
 By default the {\em reeval\_ready} field is set to {\em
   compute\_dependencies\_first}, but when {\sf
   traverse\_dependent\_nodes()} adds a subgoal $S'$ to the invalid
 list the {\em reeval\_ready} field for $S'$ is set to {\em
   compute\_directly} so that the next call to $S'$ will not add it
 again.  Later, the {\em reeval\_ready} field is reset to {\em
   compute\_dependencies\_first} in {\sf incremental\_reeval()} after
 its associated goal is re-evaluated (Fig.~\ref{alg:wfs-update}, line
 22); or it is reset when the IDG node frame's {\em falsecount} is set
 to 0 by {\sf propagate\_validity()} (this change to
 Fig.~\ref{alg:basic-update} is not shown).

The implementation of line 15 of Fig.~\ref{fig:lazy} uses a general
interrupt mechanism whereby a given goal $G$ may dynamically interrupt
the current execution environment $Env$ so that $G$ is immediately
executed and success and failure continuations of $G$ are (a
modification of) $Env$\footnote{In XSB, as in other Prologs, such
  interrupts are used to handle unification of attributed variables,
  signaling among Prolog threads, and other tasks.}.  In line 17, the
interrupt mechanism intersperses a call to {\sf
  recompute\_dependent\_tables()}, to traverse the invalid list and
recompute subgoals.  When {\sf recompute\_dependent\_tables()}
finishes, its continuation will make a fresh call $S$, which will see
a completed and valid table, and will then simply backtrack through
answers for $S$ (starting with line 18 of Fig.~\ref{fig:lazy}).

\begin{figure}[hbtp] 
\begin{footnotesize}
{\sf
\begin{tabbing}
fooo\=foo\=foo\=foo\=foo\=foo\=fooooooooo\=oooooooooooooooooo\=ooooooooooooo\=\kill
Instruction \underline{{\sf tabletry}} 
\> \> \> \> \> \> {\rm /* SLG-WAM instruction for calling a tabled subgoal $S$*/} \\ 
\> Check whether there is a table for a variant of $S$ and make a table for $S$ if not \\
\> If $S$ is incremental create an IDG node frame \\
\> \> If $S$ has a nearest tabled ancestor $S_{anc}$ add \callgr{} edges between $S$ and $S_{anc}$ if not present \\ 
5\> If there was not a table for $S$  \\ 
\> \> Create a subgoal frame for $S$ \\ 
\> \> Create a generator choice point to produce answers via program clause resolution \\ 
\> Else if $S$ is incomplete \> \> \> \> 
\> \> {\rm /* all answers for $S$ may not yet have been derived  */}\\ 
 \> \> Create a consumer choice point to perform answer resolution\\
10\> Else, if $S$ is completed \> \> \> \> \> \> {\rm /* all answers for $S$ have been derived */}\\ 
\> \> Set up a consumer choice point to perform answer resolution \\
\> \> \underline{If $S$ is incremental and invalid} \\ 
\> \> \> \> \underline{If {\em S.IDG\_node.reeval\_ready ==   compute\_dependencies\_first}} \\ 
\> \> \> \> \> \underline{{\em invalid list =  traverse\_dependent nodes(S.IDG\_node.invalid)}} \\ 
15 \> \> \> \> \> \underline{Interrupt to call {\em recompute\_dependent\_tables} with continuation $S$}\\ 
\> \> \> \> \underline{Else} \> \> \> {\rm /* {\em S.IDG\_node.reeval\_ready == compute\_directly} */} \\ 
\> \> \> \> \> \underline{{\em  incremental\_reeval(S.subgoal\_frame)}} \\ 
\> \> Branch to the instruction of the root of the answer trie for $S$\\
\> \\
\underline{traverse\_dependent\_nodes(IDG node frame {\em IDGN})} \\ 
\> For each {\em IDGN}$_{dep}$ upon which {\em IDGN} directly depends\\ 
\> \> If ({\em  IDGN$_{dep}$.reeval\_ready == compute\_dependencies\_first}) \\
\> \> \> Add {\em IDGN}$_{dep}$ to the global invalid list. \\
\> \> \> traverse\_dependent\_nodes({\em IDGN}$_{dep}$) 
\end{tabbing}
}
\end{footnotesize}
\caption{Schematic pseudo-code for lazy recomputation} \label{fig:lazy}
\end{figure}

\mycomment{ There are situations where it is convenient or necessary
  to abolish an incremental table rather than simply incrementally
  updating it.  An example of this occurs when an exception is thrown.
  If an exception is thrown over a choice point to a completed table
  no action need be taken; however if an exception is thrown over a
  choice point to an incomplete tabled subgoal $S$, XSB abolishes the
  table as its computation has become compromised.  In fact, if $S$ is
  part of a larger recursive component, the other subgoals in the
  recursive component are also compromised so that XSB allows
  exceptions to be caught only by subgoals that are ``in between''
  recursive components, an implementation choice that helps XSB to be
  stable when exceptions are thrown in heavily tabled computations.
  Consider the issues that arise when an exception or other action
  causes an incremental table for $S_{inc}$ to be abolished under
  basic incremental tabling.  This mechanism depends for correctness
  on the connectivity of the \callgr{}, if $S_{inc}$ is abolished, any
  tables that $S_{inc}$ affects must also be abolished.  Lazy
  recomputation can simply abolish $S_{inc}$ after calling {\em
    traverse\_affected\_nodes()}.  When a call is made to an affected
  node, portions of the \callgr{} that were removed through abolishing
  are reconstructed during the calls made by {\em
    incremental\_reeval()}, since the previously abolished subgoals
  are treated as new.  }
\vspace{-0.15in}

\subsection{View Consistency} \label{sec:consistency}
A fundamental principle of databases is to support view consistency:
that is, to ensure that answers to a query $Q$ should be those
derivable at the time $Q$ was begun, and should not be affected by any
updates.  Accordingly, the ISO standard for Prolog~\cite{ISO-Prolog}
specifies that an update $\upsilon$ to dynamic code should not affect
the behavior of choice points that were created before $\upsilon$.
Extending view consistency to incremental tables is critical for
understandable system behavior, especially when KRR features such as
hypothetical reasoning must be supported.  Because XSB's incremental
tabling does not allow updates that affect tables that are still being
computed (Section~\ref{sec:background}), supporting view consistency
effectively means ensuring consistency for choice points into
completed tables.  As such choice points correspond to database
cursors, we term them {\em Open Cursor Choice Points, (OCCPs)}.


The approach to view consistency adopted by \newinc{} is summarized
in this section, with further details provided in \ref{sec:app-view}.
A main goal is to avoid overhead when there are no choice points whose
``view'' needs to be maintained (including those of non-incremental
tables).  For this purpose, an {\em occp\_num} field is maintained in
the subgoal frame of a completed incremental table $T$ to indicate
whether there are OCCPs for $T$ (Appendix~\ref{app:occp-num}).  {\em
  occp\_num} is incremented when the subgoal for $T$ is called; and
decremented when the last answer for $T$ has been returned to the
call, or when a cut or throw removes the call from the choice point
stack.  Only if {\em occp\_num}$>0$ must the OCCP's view be preserved.
\Newinc{} performs this preservation during {\sf
  incremental\_reeval()} by calling {\sf preserve\_occp\_views()}
(Fig.~\ref{alg:wfs-update}, line 4).  While {\sf
  preserve\_occp\_views()} is fully described in
Appendix~\ref{app:preserve-views}, its main actions are as follows.
The choice point stack is traversed, and for each OCCP $CP_T$ for $T$,
the answer substitutions that have not yet been resolved by $CP_T$ are
determined and then copied from $T$ into the heap as a list (making
sure that their heap space is frozen so they are not lost upon
backtracking).  For each answer substitution that corresponds to a
answer whose truth value is {\em undefined}, the copying includes a
special marker {\em undef}.  Next, the structure of $CP_T$ is altered,
and its instruction is modified to backtrack through the list on the
heap rather than through the table.  Once {\sf
  preserve\_occp\_views()} has executed, {\sf incremental\_reeval()}
proceeds as it would otherwise do.  Later, when the modified version
of $CP_T$ is backtracked into, a new instruction, {\sf
  preservedViewMember} is called to return the answer substitutions
for the preserved view (Appendix~\ref{app:backtracking-pres}) using
the correct truth value.  When the answers in the list have been
exhausted, the heap space used for the list is unfrozen if it is safe
to do so.

\vspace{-0.15in}

\section{Abstracting the \Callgr} \label{sec:abs}
The \callgr{} is clearly essential to efficiently update incremental
tables, but in certain situations constructing the \callgr{} can cause
non-trivial overheads in query time and table space.  These overheads
can be addressed in many cases by {\em abstracting} the \callgr{}.
When a tabled subgoal $S$ is called, rather than creating an edge
between $S$ and its nearest tabled ancestor $S'$ (if any), one could
abstract $S$, $S'$ or both.  The semantics and implementation of
subgoal abstraction was defined in~\cite{RigS13}, here we appeal to an
intuitive notion of {\em depth abstraction}: given a subgoal $S$ and
integer $k$, subterms of $S$ with depth $k+1$ are replaced by unique
new variables.  For instance, in Fig.~\ref{fig:p-inc}, abstracting
{\em q(f(1))} at level 1 gives {\em q(f(X$_1$))}; abstracting at level
0 gives {\em q(X$_1$)}.

\begin{figure}[ht]
\centering
\begin{minipage}[b]{0.35\linewidth}
{\em
\begin{tabbing}
fooofoofoofoofoofoofoofooooooooooooooooooooooooooo\=ooooooooooooo\=\kill
:- table reach/2 as incremental.\\
:- dynamic edge/2 as incremental.\\
reach(X,Y):- edge(X,Y). \\
 reach(X,Y):- reach(X,Z),edge(Z,Y). \\
\end{tabbing}
}
\end{minipage}
\begin{minipage}[b]{0.30\linewidth}
\includegraphics[width=\textwidth]{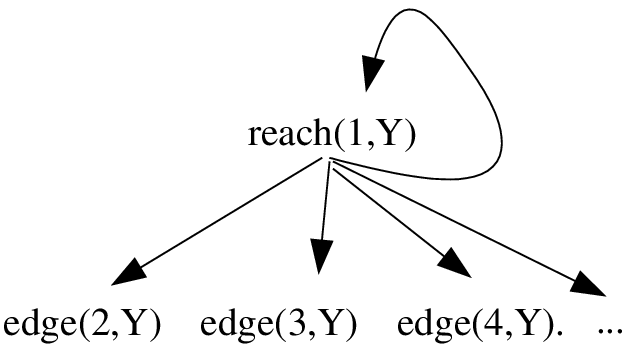}
\end{minipage}
\quad
\begin{minipage}[b]{0.30\linewidth}
\includegraphics[width=0.5\textwidth]{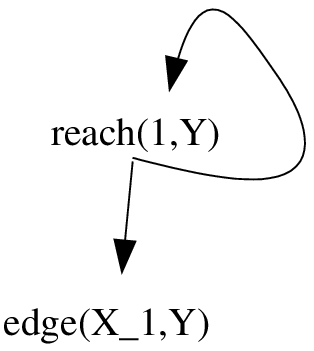}
\end{minipage}
\caption{A left-linear program and schematic \callgr{}s: Left
  without \abstraction{}; Right: with
  \abstraction} \label{fig:abstraction}
\end{figure}

Figure~\ref{fig:abstraction} illustrates an important case where
abstracting the \callgr{} can be critical to good performance for
incremental tabling.  In the case of left-linear recursion,
if no abstraction is used a new node will be created for each call to
{\em edge/2} as shown on the left side of this figure.  If a large
number of data elements are in fact reachable, the size of the
\callgr{} can be very large.  If calls to the {\em edge/2} predicate
make use of depth-0 abstraction, the graph may be much smaller as seen
on the right side of Fig.~\ref{fig:abstraction}.  Whether
abstracting a \callgr{} in this manner is useful or not is application
dependent; however, performance results in the next section illustrate
cases where abstraction greatly reduces both query time and space.

Abstracting the {\em edge/2} predicate has subtle differences from
abstracting tabled subgoals.
In the first place, the {\em edge/2} predicate of
Fig.~\ref{fig:abstraction} is not tabled.  Furthermore, the actual
{\em edge/2} subgoal itself should not be abstracted to depth 0 since
losing the first argument instantiation would prevent the use of
indexing.  Rather, only the \callgr{}'s representation of the subgoal
should be abstracted.  Fortunately, in XSB the code to intern dynamic
goals for the \callgr{} shares code used for tabling, so that
extending abstraction to handle dynamic incremental predicates is
relatively straightforward.  In XSB, abstraction of dynamic code for
the IDG can be specified via the declaration:\\
{\em \: :-~dynamic edge/2 as incremental, abstract(0)}.

\vspace{-0.15in}
\mycomment{
The \callgr{} for a set of incremental predicates can be coarsened
using XSB's implementation of subgoal abstraction~\cite{RigS13}.
Subgoal abstraction is a mechanism that allows a tabled subgoal $S$ to
be abstracted to a subgoal $S_{abs}$ such that $S_{abs}\theta = S$.
Although subgoal abstraction was originally implemented to provide
stronger termination properties for tabled evaluations, using subgoal
abstraction provides a mechanism to coarsen a \callgr{} by coalescing
different vertices for a given tabled predicate.  In \version{} of
XSB, subgoal abstraction is based on term depth, and can be declared as:

{\tt :- table p/2 as incremental,abstract($n$)}

\noindent
so that terms whose depth is greater than $n$ are abstracted: e.g.,
the level 2 term depth abstraction of {\em p(f(g(1)),g(2))} is
represented as {\em p(f(g(X$_{1.1.2}$)),g(2))}.  An important case of
this is 0-level abstraction that would abstract the above term to {\em
  p(X$_1$,X$_2$)}.  \newinc{} extends abstraction to incremental
dynamic facts, via declarations such as:

For programs that make heavy use of incremental tabling, the question
arises of how to use it efficiently.  Consider a goal $D$ to a dynamic
predicate that directly affects a set $\cS$ of tabled subgoals to a
given predicate.  The \callgr{} will contain an edge from $D$ to each
$S_i \in \cS$, and To the extent that these subgoals unify, updates
that unify with $D$ will perform redundant work as the various $S_i$
are updated.  Conversely, a \callgr{} that contains a large set $\cD$
of dynamic incremental goals, such as $p(1,Y)$, $p(2,Y)$, $p(3,Y),
\ldots$, may or may not work well compared to a \callgr{} that simply
contains $p(X,Y)$.  If the $p(n,Y)$ affect different subgoals,
maintaining the instantiation in the first argument can provide an
important mechanism to reduce the amount of reevaluation needed when a
change is made to $p/2$.  However, to the extent that the $p(n,Y)$
affect the same set of subgoals, keeping such subgoals is inefficient,
as each dynamic goal requires space for a \callgr{} leaf and edge.

The \callgr{} for a set of incremental predicates can be coarsened
using XSB's implementation of subgoal abstraction~\cite{RigS13}.
Subgoal abstraction is a mechanism that allows a tabled subgoal $S$ to
be abstracted to a subgoal $S_{abs}$ such that $S_{abs}\theta = S$.
Although subgoal abstraction was originally implemented to provide
stronger termination properties for tabled evaluations, using subgoal
abstraction provides a mechanism to coarsen a \callgr{} by coalescing
different vertices for a given tabled predicate.  In \version{} of
XSB, subgoal abstraction is based on term depth, and can be declared as:
\begin{center}
{\tt :- table p/2 as incremental,abstract($n$)}
\end{center}
\noindent
so that terms whose depth is greater than $n$ are abstracted: e.g.,
the level 2 term depth abstraction of {\em p(f(g(1)),g(2))} is
represented as {\em p(f(g(X$_{1.1.2}$)),g(2))}.  An important cass of
this is 0-level abstraction that would abstract the above term to {\em
  p(X$_1$,X$_2$)}.  \newinc{} extends abstraction to incremental
dynamic facts,  via declarations such as:
\begin{center}
{\em :- table p/2 as incremental,abstract($n$)}
\end{center}

\noindent
As will be shown in Section~\ref{sec:perf}, the use of subgoal
abstraction can have major effects on the time and space required for
incremental tabling, indicating that determining cost metrics for how
much abstraction to use for a given program and set of queries is an
important open question for incremental tabling.

}

\section{Performance Results and Analysis} \label{sec:perf}
The performance of \oldinc{} in XSB has been analyzed previously, most
extensively in \cite{Saha06}.  By and large the behavior of \oldinc{}
features are not affected by the rewriting to support \newinc{}.
Accordingly, the performance questions addressed here analyze new
features, scalability, and the behavior of incremental tabling for
KRR-style computations.  A summary of performance results is given in
this section, with tables and other details provided
in~\ref{app:performance}.

{\em Left-Linear Recursion}
Recursion is heavily used in KRR-style programs that make use of
features such as Hilog or defeasibility.  As a first test, queries of
the form $reach(\langle free \rangle, \langle free \rangle)$ were made
to a left recursive predicate (Fig.~\ref{fig:abstraction}) with and
without \abstraction{} on the {\em edge/2} predicate
(cf. Appendix~\ref{app:lr-perf}).
In the benchmarks, {\em edge/2} consists of ground facts representing
randomly generated graphs of 50,000 -- 5,000,000 edges.  As shown in
Fig.~\ref{tab:reach-over} if \abstraction{} is not used, creating
the \callgr{} adds a CPU time overhead of roughly 50\% and a table
space overhead of about 300\% compared to non-incremental tabling.  By
using \abstraction{} at depth 0, the table space overhead becomes
approximately 30\%, and the time overhead 5-10\%.
Fig.~\ref{tab:reach-upd} shows that for a batch updates
(0.02\%-2\% of EDB), the overhead of re-evaluation is negligible,
particularly if abstraction is used.

{\em Non-Stratified Linear Left Recursion}
Similar tests were made using the predicate {\em ureach/2}
(Fig.~\ref{fig:reach}), constructed to perform transitive closure, but
producing answers with truth value {\em undefined}.  The query
$ureach(\langle free \rangle, \langle free \rangle)$ was evaluated on
a graph of 500,000 edges.  Overhead results for the initial query
(Fig.~\ref{tab:ureach-over}) are similar to those for $reach(\langle
free \rangle, \langle free \rangle)$ in terms of time; however the
space overhead for incremental tabling is proportionally less (around
10-15\% with \abstraction ) as storing the conditional answers used in
this test imposes its own space overhead.  Fig.~\ref{tab:ureach-upd}
shows the time to perform various inserts that cause new answers to be
added to the table for $ureach(\langle free \rangle, \langle free
\rangle)$, and that also change the truth value of some known answers
from {\em undefined} to {\em true} as discussed in
Section~\ref{sec:wfs}.  The figure shows that updating conditional
answers imposes essentially no overhead compared to unconditional
answers.

{\em Performance Analysis on a Program with KRR Features}
The program in Fig.~\ref{fig:sn} represents a social network in
which certain members of a population are at risk, and other members
of the population may influence the behavior of the at-risk members.
While the program contains stratified negation, its main computational
challenge arises from its heavy use of equality between constants and
functional terms -- a reasoning capability similar in flavor to some
description logics.  
%
This use of equality over functional
terms quickly leads to non-termination and unsafe negative subgoals
during query evaluation.  
As discussed in Appendix~\ref{app:sn1-perf}, these behaviors are
handled using various tabling mechanisms, so that the ability to
incrementally maintain tables for queries to this program requires the
ability to update three-valued models that arise from answer
abstraction~\cite{GroS13}, tabled negation and subgoal abstraction.

As a first benchmark˜of this program, a small EDB of about 10,000
facts
was generated, and {\em good\_infuence}$(\langle bound \rangle,\langle free
\rangle)$ was queried for 200 randomly chosen values for its first
argument.  In this case, incremental tabling caused a time overhead of
about 240\% and a space overhead of 280\% -- although further
exploration of abstraction would likely reduce these numbers.  
Next, updates of substantial sizes were performed on
the EDB, and the revaluation time was computed (Fig.~\ref{tab:sn1}).
While most of these times are near the level of noise, recomputation
of several of the predicates timed out.  Analysis of these timeouts
showed that they arose because the additional facts caused a large
number of new tables to be created when the 200 queries were
re-evaluated.
 Fig.~\ref{tab:sn2} shows the times to assert or retract, plus time to
 invalidate affected subgoals via {\sf traverse\_affected\_nodes()}.
 Except for updates to {\em parent\_of\_edb/2} which directly affect
 equality, invalidation did not take a significant amount of time

{\em Scalability Analysis on a Program with KRR Features}
As a next step, the computational burden of the equality relation in the
previously mentioned program was reduced by specializing it
as discussed in Appendix~\ref{app:sn2-perf}.  Tests were performed on
EDBs from around 100,000 -- 10,000,000 facts.  As shown in
Fig.~\ref{tab:sn-scale_1}, the space and time for these computations
scales roughly linearly.  For the EDB of about 10,000,000 facts, times
were obtained for various large batch updates and for query
re-evaluation (Figs.~\ref{tab:sn-scale-2} and \ref{tab:sn-scale-3}).
Except for updates to {\em parent\_of\_edb/2}, re-evaluation time was
very low compared to initial query time (even for initial queries
using non incremental tabling), illustrating the promise of
incremental tabling for large, reactive systems.  These benchmarks
also demonstrate the scalability of this implementation, even for very
large \callgr{}s. In Fig.~\ref{tab:sn-scale_1}, the \callgr{}
contained over 750 million edges; after the update sequences mentioned
above were applied, it contained more than 1 billion edges.
\vspace{-0.15in}

\section{Discussion}
This paper has introduced \newinc{}, which improves previous versions
of incremental tabling in both semantics and efficiency.  The semantics
of lazy recomputation (Section~\ref{sec:lazy}) together with the
preservation of view consistency (Section~\ref{sec:consistency} and
\ref{sec:app-view}) guarantee that incremental tables will {\em
  always} reflect the state of the underlying knowledge base at the
time they were queried.  This view consistency takes tabled logic
programming a step closer to deductive databases, and supports
hypothetical reasoning in KRR applications.  In addition, the ability
to update 3-valued computations (Section~\ref{sec:wfs}) is necessary
when defeasibility is used over the well-founded semantics, as well as for other features such as answer abstraction.
In terms of efficiency, lazy recomputation avoids recomputing
invalidated queries until they are requeried, and \abstraction{}
(Section~\ref{sec:abs}) can significantly reduce the amount of time
and space required for queries.  The efficiency and scalability of the
resulting implementation was summarized in Section~\ref{sec:perf} and
discussed in detail in~\ref{app:performance}.  \ref{app:use} provides
further information about how to use \newinc{} in practice.

Although the major semantic issues for incremental tabling have been
addressed in this paper, KRR-style computations incur a heavy
computational burden, and the benchmark programs do show cases where
transparent incremental tabling incurs more cost than is desirable.
An important goal is to ``guarantee'' bounds for transparent
incremental tabling when used on representative KRR programs.
For instance, for constant bounds $b_i$, initial query time using
incremental tabling should never be more than $b_i$ times that of
non-incremental tabling; recomputation time should never be
significantly more than initial query time; and the space for the
\callgr{} should never be more than $b_j$ times the space of the
tables themselves.  Such bounds may be obtained through a mixture of
program analysis (some of which may itself be incremental
cf.~\cite{HPMS00}) and adaptive incremental tabling algorithms.  Even
today, incremental tabling is starting to be used to prototype
applications in stream deductive databases and event monitoring;
continued efficiency improvements should make commercial applications
in these areas possible.

\newpage
\bibliographystyle{acmtrans}
\bibliography{longstring,all}

\newpage
\appendix

\section*{Acknowledgements}
The research in this paper was partially funded by Vulcan, Inc. and
Coherent Knowledge Systems.  The author would like to thank Paulo
Moura for latex-related help, Fabrizio Riguzzi for making the
University of Ferrara server available for benchmarks, and anonymous
reviewers for their careful comments.  Finally, the author would like
to thank Michael Kifer for finding and reporting many, many bugs in
\newinc.

\section{View Consistency and Table Updates} \label{sec:app-view}
As discussed in Section~\ref{sec:consistency} the approach to
maintaining view consistency for \newinc{} has three main parts.  (1)
a count of the OCCPs for an incremental table $T$ is always
maintained.  (2) when an update affects an incremental table, the view
of an OCCP is preserved by copying its unconsumed answers onto the
heap and altering the OCCP to use the copied answers.  (3) a new
instruction returns answers from the preserved views upon
backtracking.

More than other aspects of \newinc, the details of view consistency
support rely on tabling data structures and algorithms used by XSB,
some background for which is presented here.
\begin{itemize} 
\item {\em Answer Tries} Steps 1 and 2 use the sequence of choice
  points set up when backtracking through an answer trie, the default
  data structure used by XSB to represent answers~\cite{RRSSW98}.
  Answer tries are constructed to support {\em substitution
    factoring}, so that they contain only the information used to bind
  variables in the associated subgoals, i.e., the answer substitution
  introduced in Section~\ref{sec:wfs}.  Each trie node contains an
  SLG-WAM instruction, so that returning an answer substitution
  directly corresponds to traversing a path from the root of a trie to
  a leaf, and backtracking through the trie corresponds to traversing
  the trie in a fixed depth-first order.  A choice point is created
  whenever traversing a new node that has multiple children and is
  removed when all children have been traversed (through a {\sf
    trust}-style instruction).  
\item {\em Freeze Registers} Steps 2 and 3 make use of the SLG-WAM's
  {\bf HF} (heap freeze) register, which is used to protect terms in
  the heap from being over-written when tabled computations are
  repeatedly suspended and resumed. 
\end{itemize}
While these data structures are not unique to XSB, other engines that
differ from XSB in their representation of answers or in their
implementation of suspension and resumption may implement this
approach with suitable modifications.

\subsection{Maintaining a Count of OCCPs for a Completed Incremental Table} \label{app:occp-num}
To maintain a count of OCCPs, a field called {\em occp\_num} is
added to subgoal frames.  In addition, the first choice point created
in backtracking through an answer trie, $CP_{first}$ is modified so
that it increases the OCCP number when $CP_{first}$ is created, and
decreases the number when $CP_{first}$ is removed.  Finally, any
routines that remove choice points must also be modified to reset the
{\em occp\_num}, including code that removes choice points upon
executing a cut, and when executing a throw operation.

\subsection{Preserving Views and Altering OCCPs}~\label{app:preserve-views}
In order to preserve the views of the current OCCPs for a table $T$,
{\sf incremental\_reeval()} of Fig.~\ref{alg:wfs-update} is modified
to check whether the {\em occp\_num} in the subgoal frame $S_T$ of $T$
is non-zero.  If so, {\sf preserve\_occp\_views()} is called using
$S_T$ (described at a highly schematic level in
Fig.~\ref{fig:preserve-views}).  This routine traverses the choice
point stack from top downwards until all OCCPs for $T$ have been
located\footnote{Unlike some other Prologs, XSB has a choice point
  stack separate from the local stack.  The traversal of the choice
  point stack uses the {\em previous\_top} field of choice points;
  this field was not part of the SLG-WAM design presented in
  \cite{SaSw98}, but was added to support various forms of garbage
  collection.}.  When a choice point $CP$ is encountered whose failure
continuation points to (the instruction field of) a node in the answer
trie for $T$, the process begins of copying the answers that have not
yet been consumed by $CP$.  First, an associated choice point
$CP_{root}$ must be found.  The process of backtracking through an
answer trie can create a series of trie choice points of which $CP$
will be the last in the segment due to the order of the choice point
stack traversal.  However this series of choice points will always
form a connected segment in the choice point stack, so that finding
the first choice point of the series $CP_{root}$, is relatively
simple.  Next, using $CP$ and $CP_{root}$, the unconsumed portion of
the answer trie for $T$ is traversed; each time the traversal
encounters a leaf, a pointer to the leaf is added to a list,
$Unconsumed$~\footnote{As mentioned previously, (e.g.,
  Section~\ref{sec:background}) an answer list is preserved for
  incremental tables.  While this answer list contains a pointer to
  each leaf of an answer trie, its ordering does not correspond to the
  traversal needed to obtain the unconsumed answers of an OCCP.}.
Next, a $prservedList$ is constructed on the heap, by traversing the
elements of $Unconsumed$.  Each element of the $prservedList$ contains
a binary term $ret_2(Substitution,Condition)$.  $Substitution$
represents a given answer substitution consisting of $AnsSubstSize$
terms, one for each distinct variable in the associated subgoal.  It
is repreented as a term $ret_{AnswerSubstSize}(Args)$ where each
argument corresponds to an element of the answer substitution.
$Condition$ is $null$ for unconditional answers, and points to a
special answer {\em undef} whose truth value is {\em undefined}, and
whose use is explained below.
\begin{figure}[hbtp]
\begin{center}
{\sf
\begin{tabbing}
foo\=foo\=foo\=foo\=foofoofoofooooooooooooooooooooooooooo\=ooooooooooooo\=\kill
\underline{preserve\_occp\_views(subgoal\_frame $S_T$)} \\ 
\> Traverse the choice point stack from top until $S_T.occp\_num$ OCCPs have been located \\ 
\> For each choice point $CP$ in the choice point stack \\ 
\> \> If $CP.failure\_continuation$ points into the answer trie for $S_T$ \\ 
\> \> \> Determine the root choice point, $CP_{root}$, for $CP$ \\ 
\> \> \> Construct a list of pointers, $Unconsumed$, to leaves of unconsumed answers \\ 
\> \> \> $preservedList$ =
copy\_answer\_substitutions\_to\_heap($Unconsumed$,$CP_{root}.AnsSubstSize$) \\ 
\> \> \> coalesce\_choice\_points($CP$,$CP_{root}$,$preservedList$) \\ 
\> If ({\bf HF} reg $\not=$ bottom of heap) ${\bf HF} reg = {\bf H}reg$ \\ 
\> $S_T.occp\_num$ = 0 \\ 
\> \\ 
\underline{copy\_answer\_substitutions\_to\_heap(List\_of\_trie\_leaves
  $Unconsumed$,int $AnsSubstSize$)} \\ \> For each $leaf\_ptr$ in
$Unconsumed$ \\ \> \> Create a list element with the following
information \\ \> \> Let $Ans_{heap}$ be a skeleton with argument
$ret_{AnsSubstsize}$ and $AnsSubstsize$ free variables \\ \> \>
Instantiate each argument of $Ans_{heap}$ with an element of the
answer substitution \\ \> \> If $leaf\_ptr$ corresponds to a
conditional answer \\ \> \> \> Create a non-trailed term on the heap
$ret_2(Ans_{heap},undefined\_ptr)$ \\ \> \> Else create a non-trailed
term on the heap $ret_2(Ans_{heap},null)$ \\ \> Return the head of the
List
\end{tabbing}
}
\end{center}
\caption{Schematic pseudo-code for preserving views and altering OCCPs} \label{fig:preserve-views}
\end{figure}

Once the $preservedList$ is constructed, the choice points between
$CP$ and $CP_{root}$ inclusive are coalesced via {\sf
  coalesce\_choice\_points()} into a new choice point,
$CP_{coalesced}$.  This routine is easiest to illustrate by its
results (Fig.
\ref{fig:cps-after}).  The address of $CP_{coalesced}$ is that of
$CP$, but when $CP_{coalesced}$ is backtracked into, it will restore
the engine environment as it would be if backtracking into
$CP_{root}$, and when its choices are exhausted, it will backtrack
into the choice point prior to $CP_{root}$.  Of course,
$CP_{coalesced}$ also contains a $preservedList$ field.

In Fig.~\ref{fig:cps-after} the values of $CP_{coalesced}$ come from
$CP_{root}$ rather than from $CP$, with the exception of fields
representing heap values.  In the stack-oriented backtracking used by
Prolog, the $preservedList$ can be protected by setting
$CP_{coalesced}$ to the value of the {\bf H} register after the
construction of $preservedList$.  If there is a possibility that
tabling will suspend and resume computations, {\sf
  preserve\_occp\_views()} needs to freeze the heap space containing
these answers so that the heap cells containing them will not be
overwritten.  If ({\bf HF} reg $\not=$ bottom of heap), then there is
an active tabled computation, and the heap freeze register is set to
the value of the {\bf H} register after construction of
$preservedList$.  The previous value of the {\bf HF} will be reset
using $CP_{coalesced}${\em .previous\_hfreg} once backtracking through
$preservedList$ is done.

\mycomment{
tabling use of lazy recomputation also supports a semantics for updating
the table for a subgoal $S$ while a computation has choice points set
to answers for $S$.  The subgoal frame of each table has a field
up to $S$; and is decremented whenever answers for $S$ have been
exhausted, or a cut or a thrown exception removes choice points to
$S$.  In the {\sf tabletry} instruction, lazy recomputation is only
on the answers for $S$ just as they are.  Thus all choice points for
$S$ will see the same answers, regardless of whether incremental facts
have changed, which we term the {\em cautious semantics}.  This
semantics is not ideal: for most purposes incremental tabling would
benefit from an extension of the ISO semantics in which each choice
point saw exactly the answers that were present at that call to $S$.
At the same time the cautious semantics is easy to implement and to
understand, and $S$ will be properly updated as soon as all choice
points to $S$ have been eliminated~\footnote{A Prolog flag can be set
to allow an error can be thrown if an invalid subgoal with choice
points is called.}.}

{\footnotesize
\begin{figure}[hbtp] 
\centering
\begin{tabular}{l|l|l} \cline{2-2}
$CP_{coalesced}$  & \newcpinst              & /* Failure Continuation  */                        \\ 
                         & $ereg_{root}$           & /* Top environment in stack ({\bf E} reg)*/              \\
                         & $ebreg_{root}$          & /* Environment of top choice point ({\bf EB} reg) */           \\
                         & $hreg$                  & /* Top of heap ({\bf H} reg) */                                  \\
                         & $trreg_{root}$          & /* Top of trail ({\bf TR} reg)*/                         \\
                         & $dreg_{root}$           & /* SLG-WAM delay register */                       \\
                         & $rsreg_{root}$          & /* SLG-WAM root subgoal register */                \\
                         & $previous\_cp_{root}$   & /* Pointer to previous choice point */              \\
                         & $previous\_top_{root}$  & /* Pointer to the previous top of CP stack */ \\  
                         & $answerSubstSize$ M$_{root}$+1 & /* Number of Variables in Answer Substitution */\\
                         & $answerSubst$ [M$_{root}$]      & \\
                         & :                               & \\
                         & $answerSubst$ [0]$_{root}$      & \\
                         & $preservedList$                 & \\
                         & $previous\_hfreg$               & /* Previous SLG-WAM heap freeze register */\\ \cline{2-2}
\end{tabular}
\caption{Choice point stack after coalescing}\label{fig:cps-after}
\end{figure}
}

\subsection{Backtracking through Preserved Views} \label{app:backtracking-pres}
Fig.~\ref{fig:backtrack_views} shows the new SLG-WAM instruction
that returns an answer through a preserved OCCP view when a coalesced
choice point is backtracked into.  The instruction reconstructs the
SLG-WAM state at the time of its call (except for the heap register
which was adjusted to protect the $preservedList$).  Each answer
substitution cell of the coalesced choice point is dereferenced to a
heap or local stack cell, the dereferenced cell is bound to an element
of the answer substitution, and the binding trailed.  Afterwards, the
$preservedList$ field is reset to point to the next list element if
one is present; otherwise the {\bf HF} register is set to its before
the view was preserved, and the {\bf B} register is set to the
previous choice point.

\begin{figure}[hbtp]
\begin{center}
{\sf
\begin{tabbing}
foo\=foo\=foo\=foo\=foofoofoofooooooooooooooooooo\=ooooooooooooo\=\kill
Instruction \underline{{\sf \newcpinst}} \\
\> {\em undo\_bindings({\bf B} register)} \> \> \> \> {\rm /* does not affect answers that were copied to heap */} \\
\> Restore SLG-WAM program registers \\
\> Set up pointers to access $ret_2(Substitution,Condition)$ \\
\> If ($Condition \not = null$) delay\_negatively($Condition$) \\
\> For each cell, $answerSubst[i]$, of {\bf B}$.answerSubst$ \\
\> \> Bind argument $i$ of $Substitution$ to the dereferenced value of $answerSubst[i]$ \\
\> \> \> \> and trail the binding\\
\> If ${\bf B}.preservedList$ has been consumed \\
\> \> ${\bf B} = {\bf B}.previous\_cp$ \\
\> \> {\bf HF} reg = {\bf B}.$previous\_hfreg$ \\
\> Else make ${\bf B}.preservedList$ point to the next list element
\end{tabbing}
}
\end{center}
\caption{Schematic pseudo-code for backtracking through preserved views} \label{fig:backtrack_views}
\end{figure}

\subsection{Discussion of View Consistency in \Newinc{}}
Of course, other approaches to view consistency are possible besides
the one just presented.  Before the above was implemented, answer
tries were extended to include timestamps indicating when a given
answer was valid (analogous to that of~\cite{LO87} for dynamic Prolog
code).  However, the time and space overhead of this approach was
deemed to be too high.  The actual implementation of the heap copying
approach presented here uses XSB's general tabling code as much as
possible, so that the cost to traverse tries and copy answers is
generally very low.

It should be noted that the approach to view consistency is more
closely linked to the data structures of the XSB engine than are other
features of \newinc, as view consistency interfaces with XSB's heap
and stack freezing mechanisms.

\newpage
\section{Performance Results} ~\label{app:performance}
In the benchmarks that follow, all times are measured in seconds, and
all space is measured in bytes unless otherwise specified~\footnote{
  Except for those reported in Section~\ref{app:sn2-perf}, the
  benchmarks below
were performed on a MacBook Pro, with a dual core 2.53 Ghz Intel i5
chip and 4 Gbytes of RAM.  The benchmarks for
Section~\ref{app:sn2-perf} were performed on a server at the
University of Ferrara with 3 Intel dual-core 3.47 GHz CPUs and 188
megabytes of RAM running under Fedora Linux. The default 64-bit,
single-threaded SVN repository version of XSB was used for all tests.
Benchmark programs can be obtained at {\sf
  www.cs.sunysb.edu/\~{}tswift/interpreters.html}.}.

\subsection{Transparent Incremental Tabling and Linear Left Recursion} \label{app:lr-perf}
Recursion is heavily used in KRR-style programs that make use of
features such as Hilog or defeasibility.  As a first test, queries of
the form {\em reach}$(\langle free \rangle. \langle free \rangle)$
were made to a left recursive predicate (Fig.~\ref{fig:reach}) with
and without \abstraction{} on the {\em edge/2} predicate.  As
discussed in Section~\ref{sec:abs} and shown in
Fig.~\ref{fig:abstraction}, the \callgr{} created for such a query may
differ greatly depending on whether abstraction is used.
%
In the benchmarks, {\em edge/2} consists of ground facts representing
a randomly generated graph $G(N/M)$ where $N$ is the number of
possible nodes in the graph, while $M$ is the number of directed
edges.  Because of the left recursive form of {\em reach/2} together
with its query form, the IDG nodes for {\em edge/2} are associated
with subgoals $edge(\langle free \rangle,\langle free \rangle)$ from
clause 1 of {\em reach/2}, and $edge(\langle ground \rangle,\langle
free \rangle)$ where argument 1 is instantiated by different values of
$Z$ in clause 2 of {\em reach/2}.  Using the re-evaluation strategies
described in previous sections, any update to {\em edge/2} will cause
a re-evaluation of the subgoal {\em reach}$(\langle free
\rangle,\langle free \rangle)$ so that (in this program fragment)
maintaining nodes of the form $edge(\langle ground \rangle,\langle
free \rangle)$ provides no benefit, as their dependencies will be
captured by $edge(\langle free \rangle,\langle free \rangle)$.

\begin{figure}[hbtp] 
\centering
\begin{tabular}{r| rr | rr | rr} 
Nodes      & No incr. tabling & & Incr. tabling &   & Incr. tabling & + abstraction\\
           & CPU time & Table space   & CPU time & Table space  & CPU time. & Table space             \\  \cline{1-7}
  100,000   & 0.12  &  7,663,728       & 0.21     &  21,671,136  & 0.13      & 10,273,672              \\
 1,000,000   & 2.19  & 72,121,240       & 3.43     & 211,184,888  & 2.34      & 92,746,112              \\
10,000,000   & 40.9  & 701,364,952      & 59.7     & 2,070,845,368& 41.2      & 902,048,352              \\
\end{tabular}
\caption{Overhead for \newinc{} on query evaluation of {\em
    reach($\langle$free$\rangle$,$\langle$free$\rangle$)} over
  randomly generated graphs $G(Nodes/\frac{Nodes}{2})$
} \label{tab:reach-over}
\begin{tabular}{r| rr | rr } 
  &             &            &      & \\
Nbr of asserts    & Incr. tabling              &            & Incr. tabling      & + abstraction \\
                  & Time to read/assert/inval. & Query time & Time to read/assert/inval. & Re-query time\\ \cline{1 -5}
    100           & 0.004                      & 3.53       & 0.003                      & 2.29 \\
  1,000           & 0.023                      & 3.67       & 0.022                      & 2.29 \\
 10,000           & 0.19                       & 4.20       & 0.17                       & 2.38 \\
\end{tabular}
\caption{Updates of {\em edge/2} for the query {\em
    reach($\langle$free$\rangle$,$\langle$free$\rangle$) }over a
randomly generated graph $G(1,000,000/500,000)$
} \label{tab:reach-upd}
\end{figure}

As shown in Fig.~\ref{tab:reach-over} if \abstraction{} is not used,
creating the \callgr{} adds a CPU time overhead of roughly 50\% and a
table space overhead of about 300\%.
By using \abstraction at depth 0, the table space overhead becomes
approximately 30\%, and the time overhead 5-10\%.  Regardless of
whether abstraction is used, Fig.~\ref{tab:reach-over} demonstrates
scalability for 2 orders of magnitude; the time scales log-linearly
due to the need to maintain indices.
Fig.~\ref{tab:reach-upd} shows that for a batch updates
(0.02\%-2\% of EDB), the overhead of re-evaluation is negligible,
particularly if abstraction is used.


\subsubsection{Non-Stratified Linear Left Recursion} \label{app:undef-perf}
Similar tests were made using the predicate {\em ureach/2}
(Fig.~\ref{fig:reach}).  The query {\em
  ureach($\langle$free$\rangle$,$\langle$free$\rangle$)} was
evaluated on the $G(1000000,500000)$ graph of {\em edge/2} facts, so
that all answers to the query had the truth value {\em undefined}.
Overhead results for the initial query (Fig.~\ref{tab:ureach-over})
are similar to those for {\em
  reach($\langle$free$\rangle$,$\langle$free$\rangle$) } in terms of
time; however the space overhead for incremental tabling is
proportionally less as storing conditional answers requires its own
space overhead~\cite{SaSW00}.  Fig.~\ref{tab:ureach-upd} shows the
time to add various numbers of {\em edge\_1} facts, which causes new
answers to be added to the table for {\em
  ureach($\langle$free$\rangle$,$\langle$free$\rangle$)}, and also
changes the truth value of some known answers from {\em undefined} to
{\em true} as discussed in Section~\ref{sec:wfs}.  From
Fig.~\ref{tab:ureach-upd} it can be seen that updating conditional
answers imposes essentially no overhead compared to updating unconditional
answers.

\begin{figure}[hbtp]
\centering
\begin{footnotesize}
{\em
\begin{tabbing}
fooooooooooooooooooooooooooo\=ooooooooooooo\=\kill
\> :- table ureach/2 as incremental.\\
\> :- dynamic edge/2, edge\_1/2 as incremental.\\
\> ureach(X,Y):- reach(X,Z),edge(Z,Y). \\
\> ureach(X,Y):- edge(X,Y),undefined. \\
\> ureach(X,Y):- edge\_1(X,Y). 
\end{tabbing}
}
\end{footnotesize}
\caption{Benchmark program for non-stratified left linear recursion} \label{fig:reach}
\end{figure}

\begin{figure}[hbtp] 
\centering
\begin{tabular}{r| rr | rr | rr} 
Nodes      & No incr. tabling & & Incr. tabling &   & Incr. tabling  & + abstraction\\
           & CPU time & Table space   & CPU time & Table space  & CPU time. & Table space              \\  \cline{1-7}
  100,000    & 0.14  &  21,333,304      & 0.24     & 35,540,760   & 0.15      & 24,143,168               \\
 1,000,000   & 2.30  & 208,352,144      & 3.61     & 347,416,664  & 2.42      & 228,977,672             \\
\end{tabular}
\caption{Overhead for \newinc{} on query evaluation of the
  non-stratified program {\em
    ureach($\langle$free$\rangle$,$\langle$free$\rangle$)} over
  randomly generated graphs $G(Nodes/\frac{Nodes}{2})$
} \label{tab:ureach-over} \ \\ \ \\
\begin{tabular}{r| rr | rr } 
Nbr of asserts    & Incr. tabling              &            & Incr. tabling + abstr.     & \\
                  & Time to read/assert/inval. & Query time & Time to read/assert/inval. & Re-query time\\ \cline{1-5}
    100           & 0.005                      & 3.78       & 0.004                      & 2.591 \\
  1,000           & 0.025                      & 3.83       & 0.25                       & 2.57  \\
 10,000           & 0.21                       & 3.86       & 0.22                       & 2.58  \\
\end{tabular}
\caption{Updates of {\em edge\_1/2} for the query {\em
    ureach($\langle$free$\rangle$,$\langle$free$\rangle$)} over a
  randomly generated graph $G(1000000,500000)$ }\label{tab:ureach-upd}
\end{figure}

\subsection{Analysis of Transparent Incremental Tabling on a Program with KRR-style Features} \label{app:sn1-perf}

The program in Fig.~\ref{fig:sn} represents a social network in which
certain members of a population are at risk, and other members of the
population may influence the behavior of the at-risk members.
Although the program is simplified and idealized in its content,
computationally it requires the use of some sophisticated reasoning
features.  While the program contains stratified negation, its main
computational challenge arises from its use of equality, which
provides a reasoning capability similar in flavor to some description
logics.  The predicate {\em equals/2} allows terms using the function
symbol {\em parent\_of/1} (formed from the EDB predicate {\em
  parent\_of\_edb/2}) to be considered as equal to constants
representing individuals.  

\begin{figure}[H]
\longline
\begin{center}
\begin{footnotesize}
{\em
\begin{tabbing}
fooo\=foofoofoofoofoofoofooooooooooooooooooooooooooo\=ooooooooooooo\=\kill
good\_influence(P1,P2):-
 		influences(P1,P2), \\
\> 	        sk\_not(high\_risk(P1)),sk\_not(possible\_risk(P1)), \\
\> 		(high\_risk(P2) ; possible\_risk(P2)). \\
\>		    \\
:- table high\_risk\_association/2 as incremental. \\
high\_risk\_association(Per1,Per2):- high\_risk\_contact(Per1,Per2),has\_disease(Per2). \\
high\_risk\_association(Per1,Per2):- high\_risk\_association(Per1,Per3),high\_risk\_contact(Per3,Per2). \\
\> \\
high\_risk\_contact(Per1,Per2):- may\_share\_needle(Per1,Per2). \\ 
high\_risk\_contact(Per1,Per2):- may\_have\_unprotected\_sex(Per1,Per2). \\ 
\> \\ 
:- table high\_risk/1 as incremental. \\ 
high\_risk(Per):- high\_risk\_association(Per,\_),!. \\ 
\> \\ 
:- table possible\_risk\_association/2 as incremental, answer\_abstract(3). \\
possible\_risk\_association(Per1,Per2):- might\_be\_sexual\_partner(Per1,Per2), \\
\> 				       high\_risk\_contact(Per2,\_). \\
possible\_risk\_association(Per1,Per2):- possible\_risk\_association(Per1,Per3), \\
\> 				       might\_be\_sexual\_partner(Per3,Per2). \\
 \\ 
:- table possible\_risk/1 as incremental. \\ 
possible\_risk(Per):- possible\_risk\_association(Per,\_),!. \\ 
 \\
influences(Per1,Per2):- loves(Per2,Per1). \\
influences(Per1,Per2):- works\_for(Per2,Per1). \\
influences(Per1,Per2):-  attends\_church(Per2,Church),pastor(Church,Per1). \\
influences(Per1,Per2):- lives\_at(Per1,Loc),lives\_at(Per2,Loc). \\
 \\
may\_share\_needle(Per1,Per2):- obtained\_needle(Per1,Needle,\_Loc1), returned\_needle(Per2,Needle,\_Loc2),Per1 \= Per2. \\
may\_share\_needle(Per1,Per2):- share\_needle\_report(Per1,Per2,\_Per3). \\
 \\
might\_be\_sexual\_partner(Per1,Per2):- loves(Per1,Per2),sk\_not(related(Per1,Per2)). \\
might\_be\_sexual\_partner(Per1,Per2):- sexual\_partner\_report(Per1,Per2,\_Per3).\\
 \\
:- table related/2 as incremental. \\
related(Per1,Per2):- equals(Per1,parent\_of(Per2)). \\
related(Per1,Per2):- equals(Per1,parent\_of(parent\_of(Per2))). \\
 \\
:- table loves/2 as incremental. \\
loves(X,Y):- loves(Y,X). \\
loves(X,Y):- friend(X,Y). \\
loves(X,Y):- equals(parent\_of(X),Y). \\
loves(X,Y):- grandparent\_of(X,Y). \\
 \\
:- table equals/2 as incremental, subgoal\_abstract(3). \\
equals(X,Y):- equals(Y,X). \\
equals(parent\_of(X),parent\_of(X)). \\
equals(parent\_of(X),Y):- parent\_of\_edb(X,Y). \\
equals(parent\_of(parent\_of(X)),Y):-  parent\_of(X,Z),equals(parent\_of(Z),Y). \\
 \\
father\_of(X,Y):- equals(parent\_of(X),Y),male(Y). \\
mother\_of(X,Y):- equals(parent\_of(X),Y),female(Y). \\
grandparent\_of(X,Y):- equals(parent\_of(parent\_of(X)),Y).  \\
 \\
:- dynamic friend/2, returned\_needle/3, obtained\_needle/3, share\_needle\_report/3, sexual\_partner\_report/3 as incremental. \\
:- dynamic has\_disease/1, works\_for/2, may\_have\_unprotected\_sex/2, pastor/2, parent\_of\_edb/2, lives\_at/2,attends\_church/2 \\
\>    as incremental,abstract(0). 
\end{tabbing}
}
\end{footnotesize}
\end{center}
\caption{A social network example showing KRR features} \label{fig:sn}
\end{figure}

The EDB for this program consists of 12 different dynamic predicates
as seen at the bottom of the program~\footnote{The social network
  programs and supporting data can be found at {\sf
    http://www.cs.sunysb.edu/\~{}tswift}.}.  The use of the {\em
  parent\_of/1} function within {\em equals/2} quickly leads to
non-termination and unsafe negative subgoals during query evaluation.
Unsafe negative subgoals are soundly addressed by XSB's {\em
  sk\_not/1} which skolemizes non-ground variables in an atomic
subgoal for the purpose of calling a negative subgoal.
Non-termination is addressed in two ways.  The use of subgoal
abstraction in {\em equals/2} ensures that there will be only a finite
number of tabled queries to this predicate, and in general ensures
termination for programs with finite models~\cite{RigS13}.  However,
the predicate {\em possible\_risk\_association/2}, produces an
infinite number of answers for the benchmark data set.  The use of
answer abstraction (or restraint) for this predicate, ensures sound
(but not complete) terminating query
evaluation~\cite{GroS13}~\footnote{Briefly, if an answer has an
  argument $A$ with depth greater than a given bound, $A$ is rewritten
  so that terms with depth equal to the bound are replaced by new
  variables; then the answer $A$ is assigned the truth value {\em
    undefined}}.

Thus, the ability to incrementally maintain tables for queries to this
program requires the ability to update three-valued models that arise
from answer abstraction, and to combine with tabled negation and
subgoal abstraction.  As a first benchmark test, a small EDB of about
10,000 facts about a population of 10,000 persons was generated, and
{\em good\_infuence}$(\langle bound \rangle,\langle free \rangle)$ was
queried for 200 randomly chosen values for its first argument.  If no
incremental tabling was used, the combined CPU time for these queries
averaged 1.14 seconds and table space was about 233 megabytes --- as
discussed further below the relatively large cost for this query was
almost entirely due to the use of equality.  When transparent
incremental tabling was used with no abstraction, the cost rose to
3.02 seconds, and 865 megabytes.  By applying \abstraction{} the
initial query time dropped to 2.73 seconds and 655 megabytes.  The
purpose of this sets of declarations was only to test the overhead of
\newinc{} for queries and updates: they should not necessarily be
considered to be ``optimal'' for these tests

Fig.~\ref{tab:sn1} shows times to re-evaluate the queries to {\em
  good\_influence/2} mentioned above after inserting $N$ randomly
generated facts for a given predicate (the ``Asserts'' column); and
then after retracting these inserted facts (the ``Retracts'' column).
Most of the times in Fig.~\ref{tab:sn1} are near the level of noise,
however recomputation of several of the predicates timed
out~\footnote{Timeouts, denoted Tout in Fig.~\ref{tab:sn1}, were
  triggered after one minute.  The short timeout period was to avoid
  excessive memory consumption on the laptop benchmarking machine.
  Retracts of bulk inserts could not be measured, and are designated
  as n/a.  As the population size was 10,000, 12,500 distinct facts
  could not be generated for the unary EDB predicate {\em
    has\_disease/1}.}.  Analysis of these timeouts showed that they
arose because the additional facts caused a large number of new
(sub-)tables to be created for the 200 queries.  Usually, this only
occurred after 12,500 facts were added, but for {\em
  parent\_of\_edb/2} which strongly affects goals to {\em equals/2},
the addition of 500 facts led to a timeout, while the addition of 100
facts led to a 5.57 second recomputation time.  Although the program
is not wholly monotonic, it is largely so, and computations after
retractions were always fast.  Fig.~\ref{tab:sn2} shows the times to
assert or retract plus the time taken to invalidate affected subgoals
via {\sf traverse\_affected\_nodes()}.  Except for updates to {\em
  parent\_of\_edb/2} invalidation did not take a significant amount of
time

\begin{figure}[hbtp] 
\centering
\begin{tabular}{l| rrrr | rr } 
Predicate                      & Asserts &      &       &       & Retracts &       \\
                               & 100     & 500  &  2500 & 12500\ \ & 2500      & 12500 \\ \hline
friend/2                       & 0.08    & 0.37 & 2.36  & Tout  & 0.02     & n/a  \\
returned\_needle/3             & 0.01    & 0.01 & 0.01  & 0.01  & 0.01     & 0.01 \\
obtained\_needle/3             & 0.01    & 0.01 & 0.01  & 0.02  & 0.01     & 0.01 \\
share\_needle\_report/3        & 0.03    & 0.03 & 0.13  & 0.55  & 0.01     & 0.01 \\
sexual\_partner\_report/3      & 0.01    & 0.02 & 0.12  & Tout  & 0.01     & n/a  \\ \hline 
has\_disease/1                 & 0.01    & 0.01 & 0.01  & n/a   & 0.01     & n/a  \\
works\_for/2                   & 0.01    & 0.04 & 0.42  & 1.76  & 0.01     & 0.01  \\
may\_have\_unprotected\_sex/2  & 0.03    & 0.08 & 0.12  & 0.56  & 0.02     & 0.02  \\
pastor/2                       & 0.01    & 0.01 & 0.01  & 0.01  & 0.01     & 0.01  \\
parent\_of\_edb/2                   & 5.57    & Tout & Tout  & Tout  & n/q& n/a  \\
lives\_at/2                    & 0.01    & 0.01 & 0.07  & 2.11  & 0.01     & 0.01  \\
attends\_church/2              & 0.01    & 0.01 & 0.01  & 0.01  & 0.01     & 0.01 
\end{tabular}
\caption{CPU times to re-evaluate {\em good\_influence/2} for 200
  first-argument bindings after batch updates.  The program uses
  non-specialized equality, and the EDB size is $\cO(10^4)$.  The top
  group of predicates use depth-0 \abstraction; the bottom group has
  no \abstraction.}\label{tab:sn1}
\end{figure}

\begin{figure}[hbtp] 
\centering
\begin{tabular}{l| rrrr | rr } 
Predicate                      & Asserts &      &       &       & Retracts &       \\
                               & 100     & 500  & 2500  & 12500\ \ & 2500      & 12500 \\ \hline
friend/2                       & 0.01    & 0.01 & 0.02  &          & 0.03     &   \\
returned\_needle/3             & 0.01    & 0.01 & 0.03  & 0.14     & 0.03     & 0.14 \\
obtained\_needle/3             & 0.01    & 0.01 & 0.03  & 0.15     & 0.03     & 0.20 \\
share\_needle\_report/3        & 0.01    & 0.01 & 0.02  & 0.14     & 0.03     & 0.17 \\
sexual\_partner\_report/3      & 0.01    & 0.01 & 0.03  &          & 0.03     &      \\ \hline 
has\_disease/1                 & 0.01    & 0.01 & 0.02  & n/a      & 0.02     & n/a  \\
works\_for/2                   & 0.01    & 0.01 & 0.02  & 0.10     & 0.04     & 0.16 \\
may\_have\_unprotected\_sex/2  & 0.03    & 0.08 & 0.11  & 0.52     & 0.04     & 0.16  \\
pastor/2                       & 0.01    & 0.01 & 0.02  & 0.11     & 0.03     & 0.15  \\
parent\_of\_edb/2                   & 27.8    & Tout & Tout  & Tout     & 37.2     & Tout  \\
lives\_at/2                    & 0.01    & 0.01 & 0.02  & 0.11     & 0.03     & 0.17  \\
attends\_church/2              & 0.01    & 0.01 & 0.02  & 0.11     & 0.03     & 0.16 
\end{tabular}
\caption{CPU times to apply updates and to invalidate subgoals created
  by queries to {\em good\_influence/2} for 200 first-argument bindings.  The program uses non-specialized equality, and the EDB size is
  $\cO(10^4)$.  The top group of predicates use depth-0 \abstraction;
  the bottom group has no \abstraction.}\label{tab:sn2}
\end{figure}

\subsubsection{Scalability Analysis on a Program with KRR Features} \label{app:sn2-perf}
As a next step, the equality relation in the previously mentioned
program of Fig.~\ref{fig:sn} was specialized so that it had the
form:

\begin{center}
{\em
\begin{tabbing}
fooo\=foofoofoofoofoofoofooooooooooooooooooooooooooo\=ooooooooooooo\=\kill
:- table equals/2 as incremental, subgoal\_abstract(3). \\
{\bf {\em equals(X,Y):- atomic(X),Y = parent\_of(\_),equals(Y,X).}} \\
equals(parent\_of(X),parent\_of(X)). \\
equals(parent\_of(X),Y):- parent\_of\_edb(X,Y). \\
{\bf {\em equals(parent\_of(parent\_of(X)),Y):- parent\_of\_edb(X,Z),equals(parent\_of\_(Z),Y1),Y1 = Y.}}
\end{tabbing}
}
\end{center}
\noindent
In this form, the first clause of {\em equals/2} is changed so that
symmetry is applied only if the first argument corresponds to a
nominal individual (constant), and the second argument has a
functional form.  The fourth clause is changed so that subgoals of the
form {\em equals}$(\langle bound \rangle,\langle bound \rangle)$ are
not called by this clause, but instead subgoals of the form {\em
  equals}$(\langle bound \rangle,\langle free \rangle)$ are called.
These changes, which do not affect the semantics of the program,
significantly reduce the time and space required for query evaluation,
although goals to {\em equals/2} are still computationally expensive
to update.

With this change, a series of 200 queries as described above were
tested on EDBs ranging from around 100,000--10,000,000 facts.  As
shown in Fig.~\ref{tab:sn-scale_1}, the space and time for these
computations scales roughly linearly.  For the EDB of about 10,000,000
facts, various batch updates were timed along with time to re-evaluate
queries (Figs.~\ref{tab:sn-scale-2} and \ref{tab:sn-scale-3}).
Specifically for $N = 2500, 12500, 62500$ and 312500, $N$ asserts of
each EDB predicate were performed and timed; and then the $N$ asserted
facts were retracted and timed.  Except for updates to {\em
  parent\_of\_edb/2}, re-evaluation time was low compared to initial
query time (even compared to the initial query time for
non-incremental tabling).  These benchmarks illustrate the scalability
of this implementation of \newinc{} even for very large \callgr{}s. In
Figs.~\ref{tab:sn-scale-2} and~\ref{tab:sn-scale-3}, the \callgr{}
contained over 750 million edges; after the update sequences mentioned
above were applied, it contained more than 1 billion edges.


\begin{figure}[hbtp] 
\centering
\begin{footnotesize}
\begin{tabular}{r| rrrr |r } 
EDB Size &     Query Time & Table Space  & \Callgr{} Nodes & \Callgr{} Edges & Non-incr Query Time \\ \hline
$\cO(10^5)$ &   3.9      &  0.51 Gbytes &  22,374         &   7,362,284      & 1.7  \\
$\cO(10^6)$ &  62.1      &  5.33 Gbytes &  67,106         &  78,612,966      & 24.5  \\
$\cO(10^7)$ & 679.8      & 51.56 Gbytes & 505,972         & 753,798,584      & 391.9 \\
\end{tabular}
\end{footnotesize}
\caption{CPU times to initially evaluate {\em good\_influence/2} for
  200 first-argument bindings for EDBs of various sizes.  The program
  uses specialized equality.}\label{tab:sn-scale_1}
\end{figure}


\begin{figure}[hbtp] 
\centering
\begin{footnotesize}
\begin{tabular}{l| rrrr | rrrr } 
Predicate                      & Asserts &     &       &        & Retracts &       \\
                               &  2500 & 12500 & 62500 & 312500 &  2500    & 12500\ & 62500 & 312500 \\ \hline
friend/2                       & 3.11  & 3.16  & 2.63  & 3.51   & 3.11     & 3.16   & 2.58  & 2.91 \\
returned\_needle/3             & 3.11  & 6.59  & 2.57  & 2.96   & 3.11     & 3.21   & 2.57  & 2.87 \\
obtained\_needle/3             & 3.11  & 3.16  & 2.59  & 2.65   & 3.11     & 3.16   & 2.52  & 2.52 \\
share\_needle\_report/3        & 3.12  & 3.16  & 2.52  & 2.54   & 3.11     & 3.16   & 2.52  & 2.54 \\
sexual\_partner\_report/3      & 3.12  & 3.16  & 2.52  & 2.54   & 3.11     & 3.16   & 2.52  & 2.55 \\ \hline 
has\_disease/1                 & 3.46  & 3.51  & 2.81  & 2.80   & 3.46     & 3.50   & 2.80  & 2.81 \\
works\_for/2                   & 3.14  & 3.25  & 3.34  & 4.81   & 3.11     & 3.16   & 2.52  & 2.52 \\
may\_have\_unprotected\_sex/2  & 4.34  & 4.37  & 3.51  & 3.51   & 4.33     & 4.37   & 3.51  & 3.51 \\
pastor/2                       & 3.12  & 3.16  & 3.34  & 2.51   & 3.11     & 3.16   & 2.51  & 2.52 \\
lives\_at/2                    & 3.12  & 3.16  & 2.52  & 2.58   & 3.11     & 3.16   & 2.52  & 2.52 \\
attends\_church/2              & 3.12  & 3.16  & 2.52  & 2.52   & 3.16     & 3.16   & 2.52  & 2.52 \\ 
\end{tabular}
\end{footnotesize}
\caption{CPU times to re-evaluate {\em good\_influence/2} for 200
  first-argument bindings after batch updates.  The program uses
  specialized equality, and the EDB size is $\cO(10^7)$.  The top
  group of predicates use depth-0 \abstraction; the bottom group has
  no \abstraction.}\label{tab:sn-scale-2}
\end{figure}

 
\begin{figure}[hbtp] 
\centering
\begin{footnotesize}
\begin{tabular}{l| rrrr | rrrr } 
Predicate                      & Asserts &      &       &       & Retracts &       \\
                               &  2500 & 12500 & 62500 & 312500 &  2500    & 12500\ & 62500 & 312500 \\ \hline
friend/2                       & 0.12  & 0.60  & 3.01  & 15.9   & 0.13     & 0.67   & 3.43  & 18.1 \\
returned\_needle/3             & 0.12  & 0.60  & 3.01  & 16.0   & 0.13     & 0.69   & 3.51  & 18.4 \\
obtained\_needle/3             & 0.15  & 0.74  & 3.74  & 19.5   & 0.17     & 0.83   & 4.21  & 22.0 \\
share\_needle\_report/3        & 0.12  & 0.61  & 2.99  & 15.9   & 0.11     & 0.01   & 2.98  & 15.8 \\
sexual\_partner\_report/3      & 0.12  & 0.61  & 2.99  & 16.0   & 0.11     & 0.59   & 2.98  & 15.9 \\ \hline 
has\_disease/1                 & 0.07  & 0.33  & 1.67  & 8.6    & 0.08     & 0.39   & 1.87  & 10.5 \\
works\_for/2                   & 0.12  & 0.57  & 0.42  & 16.1   & 0.0      & 0.65   & 3.34  & 18.2 \\
may\_have\_unprotected\_sex/2  & 0.34  & 1.68  & 8.45  & 43.3   & 0.13     & 1.75   & 8.87  & 45.3 \\
pastor/2                       & 0.07  & 0.33  & 1.71  & 19.5   & 0.08     & 0.39
   & 2.04  & 10.9 \\
parent\_of\_edb/2                   & 380.9 & Tout  & Tout  & Tout   & 222.6    & Tout   & Tout  & Tout \\
lives\_at/2                    & 0.11  & 0.56  & 2.82  & 14.7   & 0.14     & 0.68   & 3.45  & 18.0 \\
attends\_church/2              & 0.07  & 0.34  & 1.71  & 8.9    & 0.08     & 0.43   & 2.15  & 11.3 \\ 
\end{tabular}
\end{footnotesize}
\caption{CPU times to apply updates and to invalidate subgoals created
  by queries to {\em good\_influence/2} for 200 first-argument bindings.
  The program uses specialized equality, and the EDB size is
  $\cO(10^7)$.  The top group of predicates use depth-0 \abstraction;
  the bottom group has no \abstraction.}\label{tab:sn-scale-3}
\end{figure}

\section{A Note on Usability} \label{app:use}
The XSB manual contains information on how transparent incremental
tabling may be used in practice; however to make this paper
self-contained, we provide an outline of some usability and system
aspects.

XSB has a variety of tabling mechanisms that are used for different
purposes.  As seen from Fig.~\ref{fig:sn}, \newinc{} works properly
with subgoal abstraction and with answer abstraction; as discussed in
Section~\ref{sec:wfs}, \newinc{} works properly with well-founded
negation regardless of the tabled negation operator: for instance with
{\em sk\_not/1} in Fig.~\ref{fig:sn}, or with other XSB operators such
as {\em tnot/1}.  It also works properly with tabled attributed
variables (supporting tabled constraints).  A variety of dynamic code
may be used as a basis for \newinc{} including not only regular facts
and rules, but also facts that are interned as XSB tries.  Incremental
tables, of whatever form, may be used alongside non-incremental
tables, although special declarations must be made if an incremental table
depends on a non-incremental table.

Within the current version of XSB, \newinc{} does not yet work
properly with call subsumption, answer subsumption, hash-consed
tables, or multi-threaded tables; also, predicates that are tabled as
incremental must use static code rather than dynamic code.  Attempts
to declare a predicate using an unsupported mixture of tabling
features causes a compile-time permission error.

There are situations where it is convenient or necessary to abolish an
incremental table rather than updating it.  An example of this occurs
when an exception is thrown.  If an exception is thrown over a choice
point to a completed table no action need be taken; however if an
exception is thrown over a choice point to an incomplete tabled
subgoal (including one that is being recomputed), XSB abolishes the
table as its computation has become compromised.
In \newinc{}, abolishing an incremental table is not problematic.  If
a table $T$ is to be abolished, tables that depend on $T$ must be
invalidated before actually abolishing $T$ itself.  When a call is
made to a subgoal with an invalidated affected node, portions of the
\callgr{} that were removed through abolishing will be reconstructed
during the calls made by {\em incremental\_reeval()}, due to the
actions of lazy recomputation.

\end{document}